\documentclass[a4paper,11pt]{article}
\pdfoutput=1

\usepackage{jcappub}

\usepackage[T1]{fontenc}
\usepackage{hyperref}

\newcommand{\dd}[0]{\ensuremath{\mathrm{d}}}
\newcommand{\ee}[0]{\ensuremath{\mathrm{e}}}
\renewcommand{\vec}[1]{{\boldsymbol{#1}}}

\title{Stochastic cosmic ray sources and the TeV break in the all-electron spectrum}

\author{Philipp Mertsch}
\affiliation{Institute for Theoretical Physics and Cosmology (TTK), RWTH Aachen University, Sommerfeldstr. 16, 52074 Aachen, Germany}

\emailAdd{pmertsch@physik.rwth-aachen.de}

\abstract{
Despite significant progress over more than 100 years, no accelerator has been unambiguously identified as the source of the locally measured flux of cosmic rays. High-energy electrons and positrons are of particular importance in the search for nearby sources as radiative energy losses constrain their propagation to distances of about $1 \, \text{kpc}$ around $1 \, \text{TeV}$. At the highest energies, the spectrum is therefore dominated and shaped by only a few sources whose properties can be inferred from the fine structure of the spectrum at energies currently accessed by experiments like AMS-02, CALET, DAMPE, \textit{Fermi}-LAT, H.E.S.S. and ISS-CREAM. We present a stochastic model of the Galactic all-electron flux and evaluate its compatibility with the measurement recently presented by the H.E.S.S. collaboration. To this end, we have MC generated a large sample of the all-electron flux from an ensemble of random distributions of sources. We confirm the non-Gaussian nature of the probability density of fluxes at individual energies previously reported in analytical computations. For the first time, we also consider the \emph{correlations} between the fluxes at different energies, treating the binned spectrum as a random vector and parametrising its joint distribution with the help of a pair-copula construction. We show that the spectral break observed in the all-electron spectrum by H.E.S.S. and DAMPE is statistically compatible with a distribution of astrophysical sources like supernova remnants or pulsars, but requires a rate smaller than the canonical supernova rate. This important result provides an astrophysical interpretation of the spectrum at TeV energies and allows differentiating astrophysical source models from exotic explanations, like dark matter annihilation. We also critically assess the reliability of using catalogues of known sources to model the electron-positron flux.
}

\notoc

\begin{document}

\hfill{TTK-18-36}

\maketitle
\flushbottom

\section{Introduction}

Identifying the origin of cosmic rays (CRs) is a century old problem~\cite{Hess:1912srp}. Much evidence has been accumulated supporting the widely accepted picture for Galactic CRs: Strong shocks in supernova remnants accelerate CRs by transforming part of the kinetic energy of the supernova ejecta into CRs~\cite{1977SPhD...22..327K,1977ICRC...11..132A,Bell1978a,Bell1978b,Blandford1978}. Upon release from the shocks, CRs diffuse through the Galactic halo by resonantly interacting with the tangled Galactic magnetic fields~\cite{TheOriginofCosmicRays1964}. For a number of Galactic supernova remnants, non-thermal particle spectra are observed from radio to X-rays to gamma-rays energies~\cite{Funk:2015ena}, a few showing hints of the spectrum extending up to hundreds of TeV. Yet, despite a century of experimental campaigns and theoretical modelling, no \emph{individual} source for the CRs measured locally has unambiguously been identified.

The dominant loss processes for CR protons are Coulomb interactions and ionisation (dominant below a few tens of MeV), inelastic collisions with the interstellar gas (dominant around $\sim 100 \, \text{MeV}$) and diffusion or advection from the CR halo (dominant above a few hundred MeV). For heavier nuclei, spallations are important in a wider energy range compared to protons. These loss times are typically larger than $10^7 \, \text{yrs}$ between tens of MeV and tens of GeV, and due to the long residence times, the flux measured locally is contributed to by a large number of sources. This makes identifying individual sources difficult as they only lead to minuscule features on top of a largely featureless power law spectrum. For instance, it was recently shown~\cite{Genolini:2016hte} that the ``discrepant hardening'' observed in CR nuclei at rigidities of a few hundred GeV~\cite{2007BRASP..71..494P,Ahn:2010gv,Adriani:2011cu,Aguilar:2015ooa,Aguilar:2015ctt,Aguilar:2017hno} cannot be due to the dominance of an individual source. An alternative for identifying nearby sources that is oftentimes referred to is the measurement of CR anisotropies and the identification of a CR source in the direction of the maximum of the dipole anisotropy. Note, however, that the directional association breaks down for anisotropic diffusion~\cite{Kumar:2014dma,Mertsch:2014cua}. Even for the case of isotropic diffusion, identification of the dipole direction with the CR gradient direction is unreliable due to our ignorance of the details of the nearby turbulent magnetic field~\cite{Mertsch:2014cua}.

The situation is somewhat different for CR electrons\footnote{In the following, we will take ``electrons'' to signify electrons and positrons, as their diffusive propagation does not depend on particle charge.}. Due to their lower mass, they are subject to severe radiative losses by synchrotron emission and Inverse Compton scattering and above $\sim 10 \, \text{GeV}$, their cooling time is smaller than the timescales for diffusive and advective losses. This severely limits the distances over which electrons can propagate. It is believed that at TeV energies, at which current observations are aimed, only a few sources contribute. Thanks to the excellent precision of direct observations from AMS-02~\cite{Kounine:2012ega,Rosier-Lees:2014,Ting:2013sea,Lee:2012,Aguilar:2012,Schael:2012,Bertucci:2011,Incagli:2010zz,Battiston:2008zza}, CALET~\cite{Torii:2015lck}, DAMPE~\cite{TheDAMPE:2017dtc} and \textit{Fermi}-LAT~\cite{Abdollahi:2017nat}, we should soon be able to discern spectral features from individual sources. To date, this might be the most promising avenue for identifying the sources of local CRs. Despite contributing only $\sim 1 \, \%$ of the total CR flux, electrons therefore open the window to charged particle astronomy in the near future. (See also Refs.~\cite{Malyshev:2009tw,Cholis:2017ccs}.) In addition, since the discovery of the positron excess by PAMELA~\cite{Adriani:2008zr}, there has been a lot of interest in measurements of CR electrons and positrons for dark matter (DM) indirect searches. Developing a sound understanding of the electron-positron fluxes expected from astrophysical sources is critical in discriminating between astrophysical or DM explanations of spectral features.

Recently, experiments have made great strides towards delivering this high precision measurement of the electron-positron flux. PAMELA~\cite{Adriani:2011xv,Adriani:2013uda} and more so AMS-02~\cite{Aguilar:2014fea,Aguilar:2014mma,Aguilar:2018ons} have measured the electron and positron spectra in the GeV-TeV range with unprecedented precision. CALET~\cite{Adriani:2017efm} and DAMPE~\cite{Ambrosi:2017wek} have contributed measurements of the all-electron spectrum, that is the sum of electron and positron spectra, up to and around a TeV. While there are indications that the positron flux is cutting off around a few hundred GeV~\cite{WengICHEP2018}, the electron spectrum is surprisingly featureless up to hundreds of GeV. At even higher energies, indirect observations from Cherenkov telescopes have indicated the presence of a spectral break around a TeV, with the spectrum softening from a spectral index $\sim -3$ to $-3.7$. This was first reported a few years back by H.E.S.S.~\cite{Aharonian:2008aa,Aharonian:2009ah}. Recently, this measurement has been extended to $\sim 20 \, \text{TeV}$~\cite{KerszbergICRC2017} and was largely confirmed by DAMPE~\cite{Ambrosi:2017wek}. Despite the importance of these observations for the question of CR origin, the TeV break in the all-electron spectrum has been interpreted almost exclusively in DM scenarios, also in connection with a very narrow feature observed by DAMPE around $1.4 \, \text{TeV}$. (See Ref.~\cite{Genolini:2018jnu} for a review and collection of references.)

In the following, we will attempt an astrophysical interpretation of the break in the all-electron spectrum. Most detailed models of CR transport, including the popular \texttt{GALPROP}~\cite{Galprop1,Strong:1998pw}, \texttt{DRAGON}~\cite{Evoli:2008dv,Evoli:2016xgn}, \texttt{PICARD}~\cite{Kissmann:2014sia} and \texttt{USINE}~\cite{Maurin:2018rmm} codes, solve the equations underlying CR transport (see below) by assuming a CR source density that is smooth in position and time. Assuming power law spectra (with exponential cut-offs) for the sources, this leads to a rather smooth turn-over in the propagated spectrum, much smoother than the observed break. We know, however, that the acceleration of CRs is taking place in spatially and temporally discrete sources~\cite{Eichler:1980hw,Cowsik:1980ApJ,Fransson:1980ApJ}. At TeV energies, where the propagation distance of CR electrons (see above) becomes smaller than the mean distance between sources, the discreteness of sources becomes important and leads to features in the spectrum from individual sources.

Effects of source discreteness and the stochasticity implied for the CR spectra have been considered in the literature before. Oftentimes this is done by supplementing a smooth density for far away sources with a number of individual, nearby sources~\cite{1970ApJ...162L.181S,Atoian:1995ux,Kobayashi:2003kp,DiMauro:2014iia} with distances and ages oftentimes provided by supernova remnant (SNR)~\cite{Green:2015isa} or pulsar catalogues~\cite{Manchester:2004bp}. We will critically review this approach below. The alternative is to treat the positions and ages of sources as random variables~\cite{Strong:2001qp,Swordy:2003ds}, such that the flux from an arbitrary, individual source also becomes a random variable. As the total flux is the sum of the fluxes from individual sources, it is tempting to evaluate its statistics with the help of the central limit theorem. The expectation value for the total flux is of course just the prediction obtained by assuming a smooth source density, but as was first pointed out for the flux of CR nuclei~\cite{Lee:1979zz}, the variance of the flux is diverging. It has been suggested that this divergence could be cured by introducing a minimum distance $s_{\text{min}}$ in the computation of the variance~\cite{2006AdSpR..37.1909P,Blasi:2011fi}. However, the level of variance is very sensitive to $s_{\text{min}}$ and it is far from clear what value to adopt for $s_{\text{min}}$~\cite{Mertsch:2010fn,Bernard:2012wt}. On closer inspection it becomes apparent that the divergence is due to a long power law tail in the probability density function (PDF) for the flux from individual sources. A small number of authors have thus instead considered a generalised version of the central limit theorem~\cite{Gnedenko:1954} which applies to PDFs with power law tails and diverging second moments. This makes the problem tractable again and it turns out~\cite{Lagutin:1995rr} that the total flux is distributed following a stable distribution~\cite{Uchaikin:1999,nolan:2010}. For nuclei, this idea was developed further~\cite{Genolini:2016hte} and applied to the question of ``discrepant hardening'' (see above).

For CR electrons, local variations in the all-electron spectrum were considered~\cite{Pohl:1998ug} as a possible explanation for a diffuse gamma-ray flux harder than what would be expected from the locally measured electron flux. Other studies~\cite{Grasso:2009ma,Kawanaka:2009dk,Blasi:2010de,Kashiyama:2010ui} have considered the variations in the locally measured electron fluxes with a view to the increasing precision of observations. While most of these studies are numerical, the single-source PDF can also be computed analytically~\cite{Mertsch:2010fn}. Upon application of the generalised central limit theorem, the variation of the electron flux was quantified in terms of quantiles of a stable distribution.

The remainder of this paper is organised as follows: We will review the transport of CR electrons in the Galaxy, in particular the Green's function approach, and present the parameters that we will adopt throughout this study in Sec.~\ref{sec:transport}. Given our ignorance of the exact distances to and ages of all sources, we will introduce a statistical model in Sec.~\ref{sec:StatisticalModel}. While it would seem that no firm predictions could be made, we will make \emph{statistical} predictions for the all-electron flux. In particular, we will consider the PDF of fluxes at individual energies and their non-Gaussian statistics. For the first time, we will also quantify the correlations between different energy bins. As the form of the Green's function in the general case is very complicated, no analytical treatment seems possible at this point. Instead, we will perform Monte Carlo (MC) simulations of the CR all-electron flux and fit a semi-analytical joint distribution to a large sample of the ensemble of spectra. This joint distribution makes use of so-called pair-copula constructions which we will introduce in Sec.~\ref{sec:PairCopulae}. We give details of our MC simulation in Sec.~\ref{sec:MC}. Finally, in Sec.~\ref{sec:results}, we will put this machinery to use and quantify the likelihood of the broken power law measured by H.E.S.S. Comparing to the distribution of likelihoods observed in the MC simulations, we will show that a model with the canonical supernova rate of $2 \times 10^4 \, \text{Myr}^{-1}$ is not compatible with the H.E.S.S. flux. However, a model with a rate lower by one order of magnitude is. We will discuss these results in Sec.~\ref{sec:discussion} and critically review the alternative approach of using catalogues for predicting the local electron-positron fluxes, showing that this approach is inherently unreliable due to catalogue incompleteness. In Sec.~\ref{sec:SummaryConclusion}, we will provide a brief summary and conclude.

\section{Modelling the CR ell-electron spectrum}

\subsection{Propagation of CR electrons}
\label{sec:transport}

The transport of CR electrons is governed by the simplified CR transport equation (e.g.~\cite{Ginzburg:1990sk}),
\begin{equation}
\frac{\partial \psi}{\partial t} - \vec\nabla \cdot \left( \kappa \cdot \vec\nabla \psi \right) + \frac{\partial}{\partial E} \left( b(E) \psi \right) = q \, .
\label{eqn:TPE}
\end{equation}
Here, $\psi = \psi(\vec{r}, t, E) = \dd{}n/\dd{}E$ denotes the isotropic part of the differential CR electron density and it is related to the differential flux $\phi = (\dd^4 n)/(\dd{}E \, \dd{}A \, \dd{}t \, \dd{}\Omega) = c / (4 \pi) \psi$ and to the phase-space density $f = (\dd^6 n)/(\dd^3 p \, \dd^3 x)$ through $\psi = (4 \pi p^2 / c) f$. Note that we are constraining ourselves to relativistic electrons here.

In general, spatial diffusion is characterised by the diffusion tensor $\kappa$ which can be space-, time- and energy-dependent. Here, we constrain ourselves to isotropic diffusion and we also neglect variations of the diffusion tensor with space and time (which we justify below). This reduces the diffusion term to $\kappa \Delta \psi$, $\kappa$ now being a (scalar) diffusion coefficient. Spatial diffusion is due to resonant interactions between the charged CRs and the turbulent magnetic field~\cite{1966ApJ...146..480J,1966PhFl....9.2377K,1967PhFl...10.2620H,1970ApJ...162.1049H} and the rigidity-dependence of the diffusion coefficient depends on the power spectrum of magnetic turbulence. For a Kolmogorov-like turbulence spectrum, the diffusion coefficient has a power law dependence on rigidity (or energy $E$, as we are only considering relativistic particles with charge $|Z| = 1$ here), $\kappa(E) = \kappa_* (E/E_*)^{\delta}$.

There is another limit, complementary to the isotropic diffusion case adopted here. This is based on the expectation that turbulence in the interstellar medium is relatively weak. In the presence of a regular magnetic field, diffusion is thus strongly anisotropic with diffusion perpendicular to the magnetic field lines being strongly suppressed. We can thus effectively treat the problem as one-dimensional, with the coefficient for diffusion along the magnetic field $\kappa_{\parallel}$. Assuming again independence from time and position, the diffusion term simplifies to $\kappa_{\parallel} (\partial^2 \psi) / (\partial z^2)$. (Here, $z$ is the distance from the observer along the magnetic field line, also referred to as ``magnetic distance''.)

Electrons are subject to strong radiative losses in the interstellar medium due to synchrotron emission and Inverse Compton scattering. We call the energy loss rate $b(E) \equiv \dd{}E/ \dd{}t < 0$. In the Thomson limit, where the energy of the (virtual) photon in the electron rest frame is much lower than the electron rest mass energy, the energy dependence is $b(E) \propto E^2$,
\begin{equation}
\frac{\dd E}{\dd t} = \frac{4}{3} \sigma_T c \beta^2 \gamma^2 (U_{\text{rad}} + U_{B}) \equiv -b_* E^2 \, ,
\end{equation}
Here, $\sigma_T$ and $c$ denote the Thomson cross-section and speed of light, while $\beta$ and $\gamma$ are the relativistic speed and Lorentz factor, respectively. The energy densities of the interstellar radiation fields (ISRFs) and of the magnetic field are $U_{\text{rad}}$ and $U_{B}$, respectively. With the initial condition $E(t_0) = E_0$ this is easily integrated to
\begin{equation}
E(t) = \frac{E_0}{1 + b_* (t - t_0) E_0} \, ,
\label{eqn:E-E0}
\end{equation}
 if $b_*$ is independent of time.
 
In the general regime, with the most important Klein-Nishina corrections~\cite{Blumenthal:1970gc},
\begin{equation}
\frac{\dd E}{\dd t} = \frac{4}{3} \sigma_T c \beta^2 \gamma^2 \sum_r U_r \left( 1 - \frac{63}{10} \frac{\gamma \langle \varepsilon_r^2 \rangle}{m c^2 \langle \varepsilon_r \rangle} \right) \, .
\end{equation}
Here, $r$ labels the various components of radiation fields, that is the magnetic field for synchrotron losses and the ISRFs for Inverse Compton losses, and $U_r$ denotes their energy densities. $\langle \varepsilon_r \rangle$ and $\langle \varepsilon_r^2 \rangle$ are the mean and mean-square photon energies. We model each component of the ISRFs with a grey body of temperature $T_{\text{r}}$ and energy density $U_{\text{r}}$. Note that the energy loss rate is also position-dependent in principle.

However, with the estimates for the energy densities of the ISRFs shown in Tbl.~\ref{tbl:parameters}, we infer a cooling time $\tau_{\text{cool}} \equiv E / |\dd{}E / \dd{}t|$ of the order $3 \times 10^5 \, \text{yr}$ at $E = 1 \, \text{TeV}$. The distances that particles can diffuse over such times are therefore relatively short, $\sim \sqrt{\kappa \tau_{\text{cool}}} \simeq 0.3 \, \text{kpc}$ (for $\kappa(1 \, \text{TeV}) \simeq 10^{29} \, \text{cm}^2 \text{s}^{-1}$). We expect variations of the diffusion coefficient and of the ISRFs to be relatively small on these scales and therefore chose to ignore them altogether.

We could solve eq.~\ref{eqn:TPE} for an arbitrary source density spectrum $q = q(\vec{r}, t, E)$ if we knew the Green's function $G_{\text{free}} = G_{\text{free}}(\vec{r} - \vec{r}_0, t-t_0; E, E_0)$ which is defined as the solution to
\begin{equation}
\frac{\partial G_{\text{free}}}{\partial t} - \kappa \Delta G_{\text{free}} + \frac{\partial}{\partial E} \left( b(E) G_{\text{free}} \right) = \delta(\vec{r} - \vec{r}_0) \delta(t-t_0) \delta(E - E_0) \, .
\label{eqn:TPE_Greens}
\end{equation}
With a change of variables~\cite{1959SvA.....3...22S}, eq.~\ref{eqn:TPE_Greens} simplifies to the heat equation with its well-known Gaussian kernel and upon transforming back, we find
\begin{equation}
G_{\text{free}}(\vec{r} - \vec{r}_0, t-t_0; E, E_0) = (4 \pi \ell^2)^{-3/2} \frac{1}{| b(E) |} \exp \left[ - \frac{(\vec{r} - \vec{r_0})^2}{4 \ell^2} \right] \delta(t - t_0 - \tau) \, .
\end{equation}
Here,
\begin{equation}
\ell^2 = \ell^2(E, E_0) \equiv \int_{E_0}^{E} \dd E' \frac{\kappa(E')}{b(E')} \quad \text{and} \quad \tau = \tau(E, E_0) \equiv \int_{E_0}^E \frac{\dd E'}{b(E')} \, .
\end{equation}

We consider a cylindrical CR halo and ignore the boundary in the radial direction. The free escape boundary condition in $z$, $G(|z| = z_{\text{max}}) = 0$ can be satisfied by the method of mirror sources,
\begin{align}
G(\vec{r} - \vec{r}_0, t-t_0; E, E_0) \! = \!\!\!\! \sum_{n=-\infty}^{\infty} \!\! (-1)^n G_{\text{free}} (\vec{r} - \vec{r}_{0,n}, t-t_0; E, E_0) \\
\text{with} \quad \vec{r}_{0,n} = \{x_{0,n}, y_{0,n}, z_{0,n} \} = \{x_0, y_0, 2 n z_{\text{max}} + (-1)^n z_0 \} \, .
\end{align}
We employ \texttt{mpmath}\footnote{\url{https://github.com/fredrik-johansson/mpmath}} in evaluating the sum as a Jacobi theta function (cf., e.g.~\cite{Mertsch:2010fn}). The sum is truncated once the individual terms are smaller than the numerical machine epsilon.

In the following, we will factorise the source spectral density $q(\vec{r}, t, E)$ into a source density $\sigma(\vec{r}, t)$ and a source spectrum $Q(E)$. The spectral density $\psi_i$ from a single source at position $\vec{r}_i$ that injected that spectrum at time $t_i$ is,
\begin{align}
\psi_i
&= \int_E^{\infty} \dd E_0' \, G(\vec{r} - \vec{r}_i, t-t_i; E, E_0') Q(E_0') \\
&= (4 \pi \ell^2)^{-1} \ee^{- s_i^2 / (4 \ell_i^2) } \frac{b(E_0)}{|b(E)|} Q(E_0) \left( (4 \pi \ell_i^2)^{-1/2} \sum_{n=-\infty}^{\infty} (-1)^n \ee^{ - (z-z_{i,n})^2 / (4 \ell^2) } \right) \\
&\equiv \psi(s_i, T_i, E) \, ,
\label{eqn:Greens}
\end{align}
where $\ell^2_i = \ell^2_i(E) = \ell^2(E, E_0(E, T_i))$ and, $E_0 = E_0(E, T_i)$ is defined through $\dd{}E / \dd{}t = b(E)$ with $E(t_0) = E_0$. We have also defined the distance $s_i \equiv \sqrt{ (x-x_i)^2 + (y-y_i)^2 }$ and we call $T_i = (t - t_i)$ the age of the source. In the following, we will consider the same power law with an exponential cut-off, $Q(E) = Q_* (E / E_*)^{-\Gamma} \exp [ - E/E_{\text{cut}} ]$ for all source spectra and assume that both the sources and the observer are in the disk, $z = z_i = 0$. We stay agnostic to as what those sources are as long as they exhibit a power law spectrum with an exponential cut-off as characteristic for astrophysical sources like supernova remnants or pulsars. The other parameters we adopt are shown in Tbl.~\ref{tbl:parameters}.
\begin{table}
\centering
\begin{tabular}{l c c}
\hline
Description & Parameter & Value \\
\hline\hline
Spatial diffusion coefficient $\kappa$ at $E_*$ 	& $\kappa_*$			& $3 \times 10^{28} \, \text{cm}^2 \text{s}^{-1}$ \\
Spectral index of $\kappa$ 				& $\delta$				& $0.6$ \\
Half-height of CR halo 					& $ z_{\text{max}}$		& $4 \, \text{kpc}$ \\
\hline
Source spectrum at $E_*$ 				& $Q_*$				& determined by fit \\
Source spectral index 					& $\Gamma$			& $2.2$	\\
Cut-off momentum 						& $E_{\text{cut}}$		& $10^4 \, \text{GeV}$ and $10^5 \, \text{GeV}$ \\
\hline
Radiation field energy densities				& $\{ U_{\text{r}} \}$	& $\{ 0.224, 0.26, 0.6, 0.6, 0.1 \} \, \text{eV} \, \text{cm}^{-3}$ \\
Radiation field temperatures 				& $\{ T_{\text{r}} \}$		& $\{ 0, 2.7, 20, 5000, 20000 \} \, \text{K}$ \\
\hline
\end{tabular}
\caption{Parameters for the transport of CR electrons adopted here.}
\label{tbl:parameters}
\end{table}

\begin{figure}
\includegraphics[width=\textwidth]{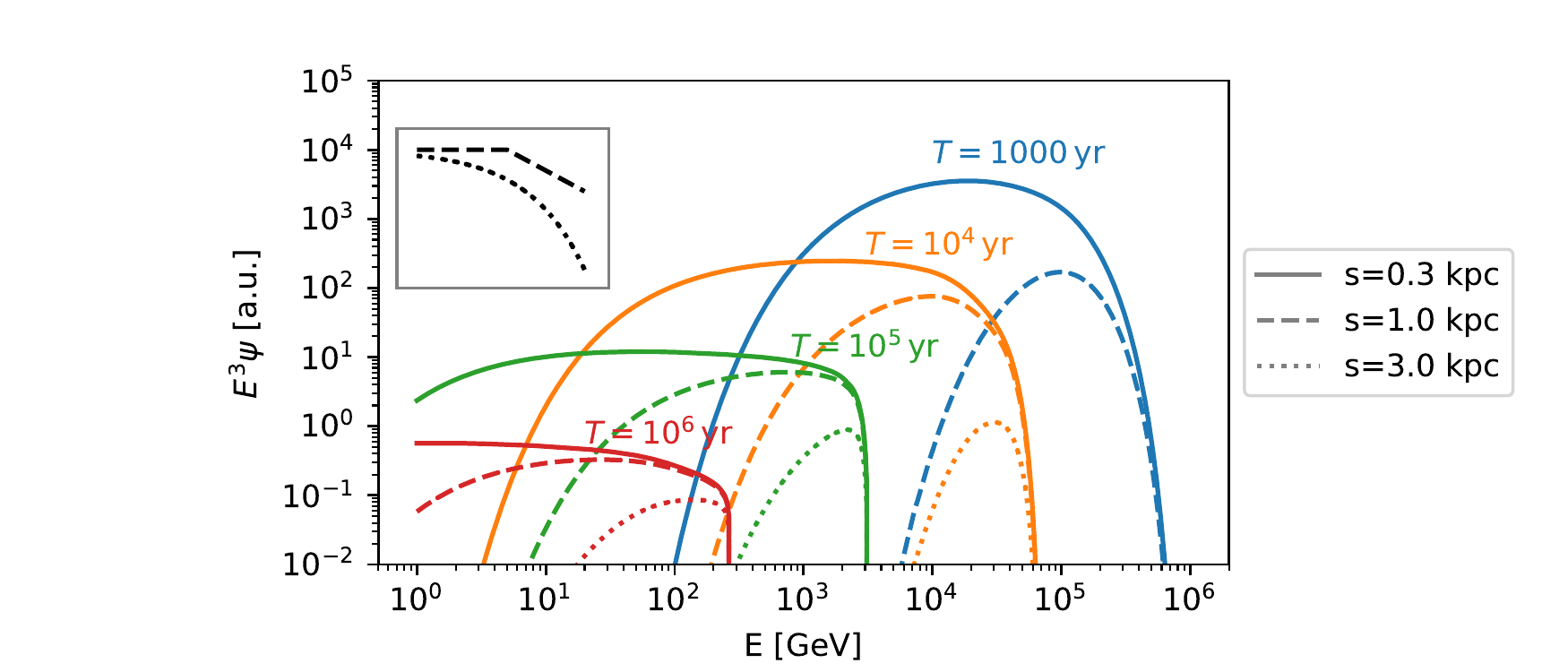}
\caption{Green's function of the simplified CR transport equation for electrons, i.e.\ eq.~\ref{eqn:Greens}, corresponding to fluxes from discrete sources in distance and age. Shown are the propagated spectra scaled with $E^3$ in arbitrary units for the three distances $s = \{0.3, 1, 3 \} \, \text{kpc}$ and the four ages $T = \{10^3, 10^4, 10^5, 10^6 \} \, \text{yr}$. For comparison, the inset shows the spectral shapes of a power law broken by one power in energy and an exponentially cut-off power law.}
\label{fig:examples1}
\end{figure}

In Fig.~\ref{fig:examples1} we present a number of example electron fluxes from combinations of the three distances $s = \{0.3, 1, 3 \} \, \text{kpc}$ and the four ages $T = \{10^3, 10^4, 10^5, 10^6 \} \, \text{yr}$. We can identify a number of effects that shape the propagated spectra: First, there is a maximum energy up to which the spectrum extends. This can be most easily seen in the Thomson approximation, cf.\ eq.~\ref{eqn:E-E0} which gives $E \to E_{\text{max}}(T) \equiv (b_* T)^{-1}$ for $E_0 \to \infty$. Note that this can lead to very sharp drops in the spectrum, in particular at late times. (To appreciate this, compare to the spectral shapes in the inset of Fig.~\ref{fig:examples1}: dashed for a power law broken by one power in energy, dotted for an exponential cut-off). Second, it takes CR electrons an energy-dependent time $\sim s^2 / \kappa(E)$ to diffuse to a distance $s$ from the source. This can be see by comparing the spectra at different distances for fixed times: At large distances, the spectrum is very peaked below $E_{\text{max}}(T)$ because low-energy CR electrons have not had time to travel to these distances. At small distances, the spectrum also extends to lower energies. Third, for a fixed distance and energy (below $E_{\text{max}}(t)$), the spectrum first increases (as discussed before) before it starts decreasing due to the ongoing diffusive spread of the particle distribution. This is similar to homogeneous and isotropic diffusion without energy losses where the density at a fixed distances is $\propto (4 \pi \kappa T)^{-3/2} \exp [ - s^2 / (4 \kappa T)]$.

\subsection{A statistical model}
\label{sec:StatisticalModel}

For a given ensemble of $N$ sources with distances $\{s_i\}$ and ages $\{ T_i \}$, $i=1, \mathellipsis, N$, the total flux is  the sum of the fluxes from individual sources,
\begin{equation}
\phi(E) = \frac{c}{4 \pi} \sum_{i=1}^N \psi(s_i, T_i, E) \, .
\label{eqn:total}
\end{equation}

In the following, we attempt a statistical model of the all-electron flux, treating the distances and ages as random variables. We thus choose to remain agnostic as to the \emph{actual} distances and ages of sources in the Galaxy and prescribe only their PDF, $\sigma(s, T)$. As these distances and ages are random variables, so will the flux from a single source and also the total flux from all the sources. As a function of energy, the local spectrum thus becomes a 1D random field, $\phi(E)$. As we will be considering fluxes at discrete energies, $\{ E_1, E_2, \mathellipsis, E_n \}$, we will be dealing with a random vector, \mbox{$\vec{\phi} = (\phi_1, \phi_2, \mathellipsis, \phi_n)^T$} where $\phi_i \equiv \phi(E_i)$.

An approach that has been followed a number of times in the literature (including a publication by these authors themselves~\cite{Ahlers:2009ae}) is to simulate a large ensemble of random realisations of source distributions, and to select the ``true'' distribution as the one that gives the best goodness of fit, oftentimes quantified by the mean-square deviation weighted with the inverse variance due to experimental errors, that is the usual $\chi^2$ test statistics. This stance is problematic in a number of ways as explained in the following.

We will assume that the propagation model outlined in Sec.~\ref{sec:transport} is exactly realised in nature. Our model is then simply characterised by the entirety of its parameters, that is the distances and ages of the sources as well as all the transport and source spectral parameters. In principle, that is in the absence of experimental errors, we should be able to achieve an arbitrarily small mean-square deviation by adopting a set of parameters arbitrarily close to the ``true'' ones. Note, however, that there will be a large degree of degeneracy in the problem, not just among the transport and source spectral parameters (e.g.\ Tbl.~\ref{tbl:parameters}) or between those and the source distances and ages. Even if we were to fix the transport and source spectral parameters to their true values, there would still be a large degree of degeneracy among the source distances and ages: Two different distributions can give total fluxes that can be arbitrarily close to each other. In turn this means that a distribution that gives a good fit need not be the true distribution or necessarily close to true distribution by some measure.

Even worse, if we were to fix the transport and source spectral parameters to the wrong values, we could still find a good fit. This is easy to see in a rather extreme example: Imagine the true sources to be homogeneously distributed with a constant source rate and soft spectra. We now try to fit their total flux by adding up the contributions from sources with very hard spectra. The propagated spectra from individual \emph{old} source will look like delta functions, with the position in energy determined by the age and the normalisation by the distribution of distances. In this manner, an arbitrary total flux can be reproduced by adding up a distribution of such delta function-like spectra in just the right way.

Furthermore, taking into account the presence of experimental statistical errors, it is not necessarily the realisation with the smallest $\chi^2$ that we should consider, but the one with a reduced $\chi^2$ around $1$. If the $\chi^2$ per degree of freedom was much smaller there is the danger of overfitting, meaning that we are exploiting the freedom in the source distribution to reduce residuals below what is to be expected for statistical errors.

Suppose we were to find a certain fraction of realisations with a reduced $\chi^2$ close to $1$. What would that mean? Would this constitute an acceptable model? (Note that what fraction of realisations is acceptable will depend on the level of experimental errors.) It appears that what this approach is lacking is a statistical estimate of how likely an acceptable $\chi^2$ should be for a given PDF of distances and ages. We will attempt at providing a prescription for assessing the statistical compatibility between the data and our model, but we will adopt a somewhat different approach. 

Considering the random vector of fluxes introduced above, \mbox{$\vec{\phi} = (\phi_1, \phi_2, \mathellipsis, \phi_n)^T$}, we will provide a parametrisation of its joint PDF, $f(\phi_1, \phi_2, \mathellipsis, \phi_n)$ for a given source PDF $\sigma(s, T)$, and with a fixed set of the transport and source spectral parameters. The joint flux PDF lends itself to a number of interesting applications:
\begin{itemize}
\item Given a measurement $\vec{\hat{\phi}} = (\hat{\phi}_1, \hat{\phi}_2, \mathellipsis, \hat{\phi}_n)^T$, we can evaluate its likelihood by substituting it into the joint flux PDF. Together with the distribution of likelihoods expected in this model, this allows us to make a statistical verdict on the compatibility of data and model.
\item We can use the joint density to extrapolate within the current model from a measurement over a limited energy range to higher energies. Imagine we have measured the spectrum in $m$ energy bins up to a maximum energy $E_m$. To evaluate the likelihood of the spectrum at higher energies $E_{m+1}, \mathellipsis, E_n$, we employ the conditional density which is the PDF with $\{ \phi_1, \phi_2, \mathellipsis, \phi_m \}$ evaluated at the measured $\{ \hat{\phi}_1, \hat{\phi}_2, \mathellipsis, \hat{\phi}_m \}$.
\item Finally, a parametrisation of the joint PDF can be useful when comparing the results from the above mentioned CR propagation codes to data. As the codes usually consider smooth source densities, they do not take into account the stochasticity of sources. At high energies where the stochasticity effects become important, the results from the codes cannot be compared to the data. A parametrisation of the joint PDF allows taking these effects into account without the need to run computationally expensive MC simulations.
\end{itemize}

In the following section, we will introduce copulae and pair-copula constructions to achieve a fast, yet flexible parametrisation of the joint PDF.

\subsection{Pair-copula construction}
\label{sec:PairCopulae}

It might be tempting to try modelling the joint flux PDF with a multivariate Gaussian. The true joint PDF cannot be a Gaussian though since a multi-variate Gaussian distribution has also (multi-variate) Gaussian marginals. However, as discussed above, the 1D marginals of the electron spectrum, that is the 1D PDFs of the flux at individual energies, are not Gaussians but stable distributions. Therefore, the joint PDF cannot be a multivariate Gaussian.

An alternative might be to use non-parametric density estimation technique. A popular method (in lower dimensions) is kernel density estimation (KDE). 
To apply this to the problem at hand, we would compute a sample from the random ensemble in a MC approach and then use this sample to construct the kernel density estimate. In higher dimensions, however, this method suffers from the curse of dimensionality, meaning that the number of sample points necessary to reliably estimate the PDF increases exponentially.

In the following we employ a technique known as pair-copula construction, a parametric, yet very flexible method for constructing multi-variate density functions. Copulae and pair-copula constructions have gained in popularity over the last few years in quantitative finance, meteorology, genetics and other fields where it is oftentimes difficult to derive parametric multi-variate densities from first principles and where multi-variate Gaussian densities do not always accurately describe the data. At the heart is the decomposition of the PDF into marginal densities and dependence structure, the latter being encoded in a \textit{copula}. This is made possible due to Sklar's theorem as we will explain in the following.

Consider a random vector $\vec{\phi} = ( \phi_1, \phi_2, \mathellipsis, \phi_n )^T$ that is distributed according to the joint probability density $f(\phi_1, \phi_2, \mathellipsis, \phi_n)$. We will also need the joint cumulative distribution function (CDF) $F(\phi_1, \phi_2, \mathellipsis, \phi_n)$ as well as the 1D marginal PDFs and CDFs, $f_i(\phi_i)$ and $F_i(\phi_i)$. Sklar's theorem~\cite{Sklar:1959} states that the joint CDF $F(\phi_1, \phi_2, \mathellipsis, \phi_n)$ can be expressed in terms of a (multi-variate) function of the 1D CDFs, called the \textit{copula} $C_{1 \mathellipsis n}$,
\begin{equation}
F(\phi_1, \phi_2, \mathellipsis, \phi_n) = C_{1 \mathellipsis n}(F_1(\phi_1), F_2(\phi_2), \mathellipsis, F_n(\phi_n)) \, ,
\end{equation}
and that $C$ is unique for continuous $F_i$.

With the $\phi_i$ being random variables, samples from the 1D CDF are also random variables and we will refer to those transformed random variables also as $F_i$. By definition, the $F_i$ are uniformly distributed with $0 \leq F_i \leq 1$, thus the copula $C_{1 \mathellipsis n}$ has uniform marginal CDFs,
\begin{equation}
C_{1 \mathellipsis n}(1, \mathellipsis, 1, F_i, 1, \mathellipsis, 1) = F_i \, .
\end{equation}

The joint density function $f(\phi_1, \phi_2, \mathellipsis, \phi_n)$ can be expressed as a product of the 1D marginals and the \textit{copula density} $c_{1 \mathellipsis n}$, i.e.\ the derivative of $C_{1 \mathellipsis n}$ with respect to $\phi_1, \phi_2, \mathellipsis, \phi_n$,
\begin{equation}
f(\phi_1, \phi_2, \mathellipsis, \phi_n) = f_1(\phi_1) f_2(\phi_2) \mathellipsis f_n(\phi_n) c_{1 \mathellipsis n}(F_1(\phi_1), F_2(\phi_2), \mathellipsis, F_n(\phi_n)) \, .
\label{eqn:marginals_copula}
\end{equation}

What has been gained by this is the separation of the behaviour of individual variables and their dependence structure. For example, an $n$-dimensional Gaussian PDF can be decomposed into the product of $n$ 1D Gaussian functions and an $n$-dimensional Gaussian copula. We can, however, also multiply the Gaussian copula with any other combination of marginal distributions, e.g.\ stable distributions, thus keeping the dependence structure, but altering the marginal behaviour.

Unfortunately, there is only a limited number of parametrisations of multivariate copulas for dimensions larger than two. However, a multivariate copula density can be constructed as a product of bivariate copula densities, a method known as \emph{pair-copula construction}. Conveniently, there is a number of widely used bivariate copulae available. Among the most common ones are the Gaussian, Clayton, Frank, Gumbel and Student's t copulae.

The factorisation into bivariate copulae builds on the factorisation of multi-variate densities into conditional densities and densities of lower dimensionality. Here, we follow Ref.~\cite{Aas:2009} in exposition and notation. We start by writing the joint density $f(\phi_1, \phi_2, \mathellipsis, \phi_n)$ as
\begin{equation}
f(\phi_1, \phi_2, \mathellipsis, \phi_n) = f_n(\phi_n) f(\phi_{n-1} | \phi_n) f(\phi_{n-2} | \phi_{n-1}, \phi_n) \mathellipsis f(\phi_1 | \phi_2, \mathellipsis, \phi_{n-1}, \phi_n) \, .
\label{eqn:factorisation}
\end{equation}
For the bivariate case ($n=2$), it follows from eqs.~\ref{eqn:marginals_copula} and \ref{eqn:factorisation}, that
\begin{equation}
f(\phi_1 | \phi_2) =  f_1(\phi_1) c_{12}(F_1(\phi_1), F_2(\phi_2)) \, .
\end{equation}

Repeating the same steps, but conditioning on $\phi_3$ gives a generalisation to three random variables,
\begin{align}
f(\phi_1 | \phi_2, \phi_3) &= c_{12 | 3}(F_{1|3}(\phi_{1} | \phi_{3}), F_{2|3}(\phi_{2} | \phi_{3})) f(\phi_1 | \phi_3) \\
&= c_{12 | 3}(F_{1|3}(\phi_{1} | \phi_{3}), F_{2|3}(\phi_{2} | \phi_{3})) c_{13}(F_1(\phi_1), F_3(\phi_3)) f_1(\phi_1) \, .
\end{align}
Here, the bivariate copula $c_{12 | 3}$ and its arguments are also conditioned on $\phi_3$. Note that an alternative decomposition would be
\begin{align}
f(\phi_1 | \phi_2, \phi_3) &= c_{13 | 2}(F_{1|2}(\phi_1 | \phi_2), F_{3|2}(\phi_3 | \phi_2)) f(\phi_1 | \phi_2) \\
&= c_{13 | 2}(F_{1|2}(\phi_1 | \phi_2), F_{3|2}(\phi_3 | \phi_2)) c_{12}(F_1(\phi_1), F_2(\phi_2)) f_1(\phi_1) \, .
\end{align}

In general, such decompositions can be written as
\begin{equation}
f(\phi | \vec{v}) = c_{\phi v_j | \vec{v}_{-j}} \left( F(\phi | \vec{v}_{-j}), F(v_j | \vec{v}_{-j}) \right) f(\phi | \vec{v}_{-j}) \, , 
\label{eqn:recursion_f}
\end{equation}
with $v_j$ one arbitrary component of the vector $\vec{v}$ of random variables, and $\vec{v}_{-j}$ is the vector with this component $j$ omitted. By repeatedly applying eq.~\ref{eqn:recursion_f} to eq.~\ref{eqn:factorisation}, we can recursively express all the conditional densities in eq.~\ref{eqn:factorisation} in terms of bivariate (conditional) copulae and marginals, thus achieving the separation of eq.~\ref{eqn:marginals_copula}. Note that in general the arguments of the copula densities are conditional distributions for which a similar recursion relation exists~\cite{Joe:1996}, that is for every $j$,
\begin{equation}
F(\phi | \vec{v}) = \frac{\partial C_{\phi, v_j | \vec{v}_{-j}} \left( F(\phi | \vec{v}_{-j}), F(v_j | \vec{v}_{-j}) \right)}{\partial F(v_j | \vec{v}_{-j})} \, .
\label{eqn:recursion_F}
\end{equation}
Eventually, we will require the form for univariate $v$ which commonly defines the so-called $h$-function,
\begin{equation}
h(\phi, v) \equiv F(\phi | v) = \frac{\partial C_{\phi, v} (\phi, v)}{\partial v} \, ,
\label{eqn:def_h}
\end{equation}
where we have specified to uniformly distributed $\phi$ and $v$, $F_{\phi}(\phi) = \phi$ and $F_v(v) = v$.

The above freedom in decomposition follows from the freedom of choice of conditioning variables in eqs.~\ref{eqn:recursion_f} and \ref{eqn:recursion_F}. It has proven useful to graphically represented the decomposition by a so-called \textit{regular vine}~\cite{Bedford:2001,Bedford:2002}, that is a series of trees, with the edges of one level forming the nodes of the next level. The nodes of the first level are just the $F_i$, that is the transformed random variables. Note that two nodes in a tree can only be connected if the corresponding edges on the previous tree level share a node.

Some vine structures may be more economical for certain dependence structures than others. For instance, in a \textit{canonical vine} or \textit{C-vine}~\cite{Kurowicka:2005}, there is one node on each tree level that connects to all the other nodes. Such a structure can be adequate if there is a hierarchy in the dependence structure of the random variables. For the problem at hand, however, we expect no such hierarchy. Instead, we expect that neighbouring random variables (that is fluxes in neighbouring energy bins) are most strongly correlated. In this case, a \textit{D-vine}, in which only neighbouring nodes are connected by edges, will be most economical. We stress that in principle all decompositions give the same multivariate copula, but certain decompositions can be more economical in the sense that the dependence structure can be localised on fewer tree levels. We refer the interested reader to Ref.~\cite{Aas:2009} for plots of example C-vines and D-vines.

The $n$-dimensional copula density is the product of $n (n-1)/2$ bivariate copulae, obtained by substituting eq.~\ref{eqn:recursion_f} into eq.~\ref{eqn:factorisation} and isolating the copula density from eq.~\ref{eqn:marginals_copula}. For the D-vine, we always pick the first conditioning variable as $v_j$ in eqs.~\ref{eqn:recursion_f} and \ref{eqn:recursion_F}. We then find for the logarithm of the copula density for the D-vine~\cite{Aas:2009},
\begin{align}
& \ln c_{1 \mathellipsis n}(F_1, F_2, \mathellipsis, F_n) \\
&= \sum_{j=1}^{n-1} \sum_{i=1}^{n-j} \ln \left[ c_{i,i+j | i+1, \mathellipsis, i+j-1} \left( F(\phi_i | \phi_{i+1}, \mathellipsis, \phi_{i+j-1}), F(\phi_{i+j} | \phi_{i+1}, \mathellipsis, \phi_{i+j-1}) \right) \right] \, .
\label{eqn:likelihood}
\end{align}

All of the bivariate copulae mentioned above have one parameter and are symmetric, that is invariant under exchange of their variables. Therefore, for an $n$-dimensional random vector, the $n$-dimensional copula possesses $n (n-1)/2$ free parameters. For the Gaussian copula, this is the same number as the number of independent off-diagonal elements of the correlation matrix, to which the pair-copula parameters can be related. We have considered a number of pair-copula families, but focussed on the Gaussian copula, the Gumbel and the Clayton survival copula after inspection of 2D scatter plots and histograms of transformed random variables from the MC simulation. In appendix~\ref{sec:appendix}, we list the copula densities and $h$-functions of these three copula families for reference and show contour plots.

With a pair-copula construction as described above, nothing has been gained in terms of evading the curse of dimensionality. The conditional pair-copulae, cf. eq.~\ref{eqn:recursion_f} will in general depend on the conditioning variables, e.g.\ through their pair-copula parameters. In the absence of an \textit{a priori} prediction for this dependence, we would again need an exponentially large sample to fit this with a non-parametric method. Instead, it has proven useful~\cite{Nagler:2016} to make the simplifying assumption that the pair-copula parameters are independent of the conditioning variables. Note that for the Gaussian copula, this assumption is actually exact. The resulting pair copula constructions are known as \emph{simplified vine models}.

The log-likelihood eq.~\ref{eqn:likelihood} then simplifies to
\begin{align}
& \ln c_{1 \mathellipsis n}(F_1, F_2, \mathellipsis, F_n) \\
&= \sum_{j=1}^{n-1} \sum_{i=1}^{n-j} \ln \left[ c_{i,i+j} \left( F(\phi_i | \phi_{i+1}, \mathellipsis, \phi_{i+j-1}), F(\phi_{i+j} | \phi_{i+1}, \mathellipsis, \phi_{i+j-1}) \right) \right] \, .
\end{align}
(Note the omission of the conditioning variables in the density $c_{i,i+j}$.) In the following, we adopt exactly this simplifying assumption for all our pair copulae.

Having fixed the structure of the pair copula construction, we now turn to the inference of its parameters. We determine those by fitting our pair-copula construction to a large sample of random vectors, $\{ \vec{\phi}_s = (\phi_{1,s}, \mathellipsis, \phi_{n,s})^T \}$ from our MC simulation, that is we find the set of pair copula parameters, that maximises the copula density. To this end, we need the transformed random variables, $\{ \vec{F}_s = (F_1(\phi_{1,s}), \mathellipsis F_n(\phi_{n,s}) )^T \}$. As we do not have a parametric form for the CDFs $F_i$, we approximate them with the \textit{empirical cumulative distribution function} $\hat{F}_i$,
\begin{equation}
F_i(t) \approx \hat{F}_i(t) = \frac{1}{N_s} \sum_{s=1}^{N_s} I(\phi_{i,s} < t) \, .
\label{eqn:eCDF}
\end{equation}
Here, $\phi_{i,s}$ is the $s$-th sample of $\phi_i$ (the flux in energy bin $i$), $N_s$ is the number of samples and $I(\cdot)$ denotes the indictor function. 

This completes our description of the construction of the joint density by a pair-copula method. In Sec.~\ref{sec:results}, we will apply it to quantify the likelihood for the stochastic source model to produce the broken-power law spectrum as observed by H.E.S.S. or DAMPE. We compute this likelihood by evaluating the joint density $f(\vec{\phi})$, see eq.~\ref{eqn:marginals_copula}, with the measured flux, $\vec{\phi} = \vec{\hat{\phi}} = ( \hat{\phi}(E_1), \mathellipsis \hat{\phi}(E_n) )^T$ where $\hat{\phi}(E)$ is the broken power law fit presented by the H.E.S.S. collaboration~\cite{KerszbergICRC2017},
\begin{equation}
E^3 \hat{\phi}(E) = \phi_0 \left( \frac{E}{1 \, \text{TeV}} \right)^{3-\Gamma_1} \left( 1 + \left( \frac{E}{E_b} \right)^{1/\alpha} \right)^{-\alpha(\Gamma_2 - \Gamma_1)} \, ,
\label{eqn:HESS_broken_power_law}
\end{equation}
with $\phi_0 = (104 \pm 1) \, \text{GeV}^2 \, \text{m}^{-2} \, \text{s}^{-1} \, \text{sr}^{-1}$, $\Gamma_1 = 3.04 \pm 0.01$, $\Gamma_2 = 3.78 \pm 0.02$, $E_{\text{b}} = 0.94 \pm 0.02 \, \text{TeV}$ and $\alpha = 0.12 \pm 0.01$.

\subsection{Details on the Monte Carlo simulations}
\label{sec:MC}

The sample from the random ensemble of all-electron fluxes that is used to fit the pair copula construction is generated as follows: We consider a source density $\sigma(s, T)$ (see Sec.~\ref{sec:transport}) that factorises into a source rate $\mathcal{R}$, which we take to be constant in time, and a spatial PDF. In galacto-centric coordinates, the spatial PDF is a logarithmic spiral with four arms of pitch angle $12.6^{\circ}$~\cite{Vallee:2005} and also contains a central bar of length $6 \, \text{kpc}$ inclined at $30^{\circ}$ with respect to the direction Sun-Galactic Centre. We have normalised it such that the radial distribution reproduces the one inferred for supernova remnants~\cite{Case:1998qg}. (See Ref.~\cite{Ahlers:2009ae} for plots of the 2D spatial density and the 1D PDF of distances $s$.) We have only considered distances up to a maximum $s_{\text{max}}$ which we fixed to $10 \, \text{kpc}$ as sources farther away contribute very little. The minimum energy of $E_{\text{min}} = 10 \, \text{GeV}$ is determining the maximum age that we need to consider. In particular in the Thomson limit, $t_{\text{max}} \simeq (b_* E_{\text{min}})^{-1} \simeq 100 \, \text{Myr}$. In turn, this sets the number of sources we need to consider, $N_{\text{src}} = \mathcal{R} t_{\text{max}}$. Note that with the above spatial PDF, only about half of the Galactic sources lie within $10 \, \text{kpc}$ from the observer. We draw $10^4$ realisations of the distribution of $N_{\text{src}}$ sources, compute their fluxes with eq.~\ref{eqn:Greens} and add them up to yield $10^4$ realisations of the total all-electron flux. The transport and source spectral parameters are shown in Tbl.~\ref{tbl:parameters}. For the source rate $\mathcal{R}$, we will show results for two examples, the canonical SN rate, $\mathcal{R}_{\text{SN}} = 2 \times 10^4 \, \text{Myr}^{-1}$~\cite{1991ARA&A..29..363V,Diehl:2006cf} and a rate reduced by one order of magnitude. The source normalisation $Q_*$ does not affect the copula or the shape of the marginal PDFs, and thus we do not need to fix it before we compare to the data.

Note that the heat equation is violating causality for very short times in the sense that the Green's function does not vanish for $|\vec{r} - \vec{r}_0| > c (t-t_0)$, where $c$ is the speed of light. This can in principle be overcome by alternatives to the heat equation that correctly model the ballistic regime and its transition to the diffusive regime, but we here follow the approach of Ref.~\cite{Genolini:2016hte} that suggests to remove sources with $s_i > c \, T_i$.

\section{Results}
\label{sec:results}

We have fit pair-copula constructions based on the Gaussian, the Gumbel and the Clayton survival bivariate copulae to the sample $\{ \vec{F}_s \}$ from the Monte Carlo simulations. For both source rates $\mathcal{R}$ considered below, we have found the Gaussian pair-copula to give the best description of the MC sample based on the likelihood, cf.\ eq.~\ref{eqn:likelihood}. In the following, we therefore only report the results from the Gaussian pair-copula construction.

\subsection{Canonical supernova rate}

\begin{figure}
\centering
\includegraphics[width=0.8\textwidth]{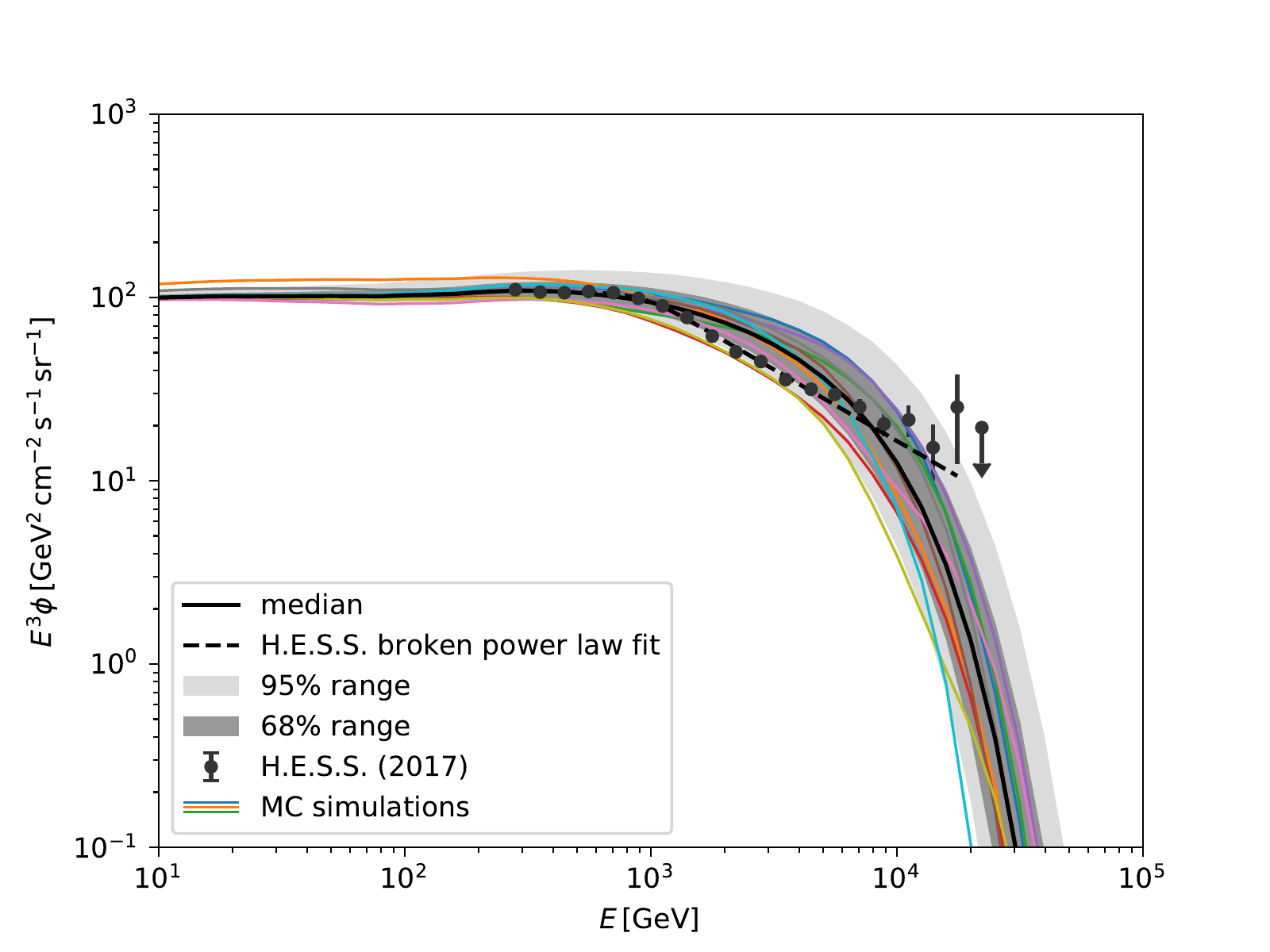}
\caption{The all-electron spectrum. The colored lines show different random realisations from the MC simulations for a Galactic source rate of $2 \times 10^4 \, \text{Myr}^{-1}$. (See Tbl.~\ref{tbl:parameters} for the other parameters used.) The black line denotes the median and the dark and light grey bands indicate the $68 \, \%$ and $95 \, \%$ variation. Finally, the data points are the measurements of the electron spectrum by H.E.S.S. and the dashed line shows the broken power law fit~\cite{KerszbergICRC2017}.}
\label{fig:spec_1e4}
\end{figure}

We start by considering the canonical supernova rate of $2 \times 10^4 \, \text{Myr}^{-1}$~\cite{1991ARA&A..29..363V,Diehl:2006cf} as source rate $\mathcal{R}$ and use the cut-off energy $E_{\text{cut}} = 10^4 \, \text{GeV}$. In Fig.~\ref{fig:spec_1e4}, a small selection of the sample of all-electron spectra from the MC computation is shown as coloured lines. We also show the median as the solid black line as well as the $68 \, \%$ and $95 \, \%$ bands around it by the dark grey and light grey bands, respectively. At the lowest energies, the scatter around the median is relatively small, but the distribution is markedly asymmetric with a long tail towards higher fluxes. In contrast, the scatter becomes larger at the highest energies, but the scatter around the median becomes more symmetric in logarithmic flux units. Note that the median is turning over rather smoothly, but individual spectra can show power law behaviour over extended energy ranges. This is due to a fortunate distribution of sources. To evaluate how fortunate, we now use the statistical machinery developed above.

\begin{figure}
\centering
\includegraphics[width=\textwidth]{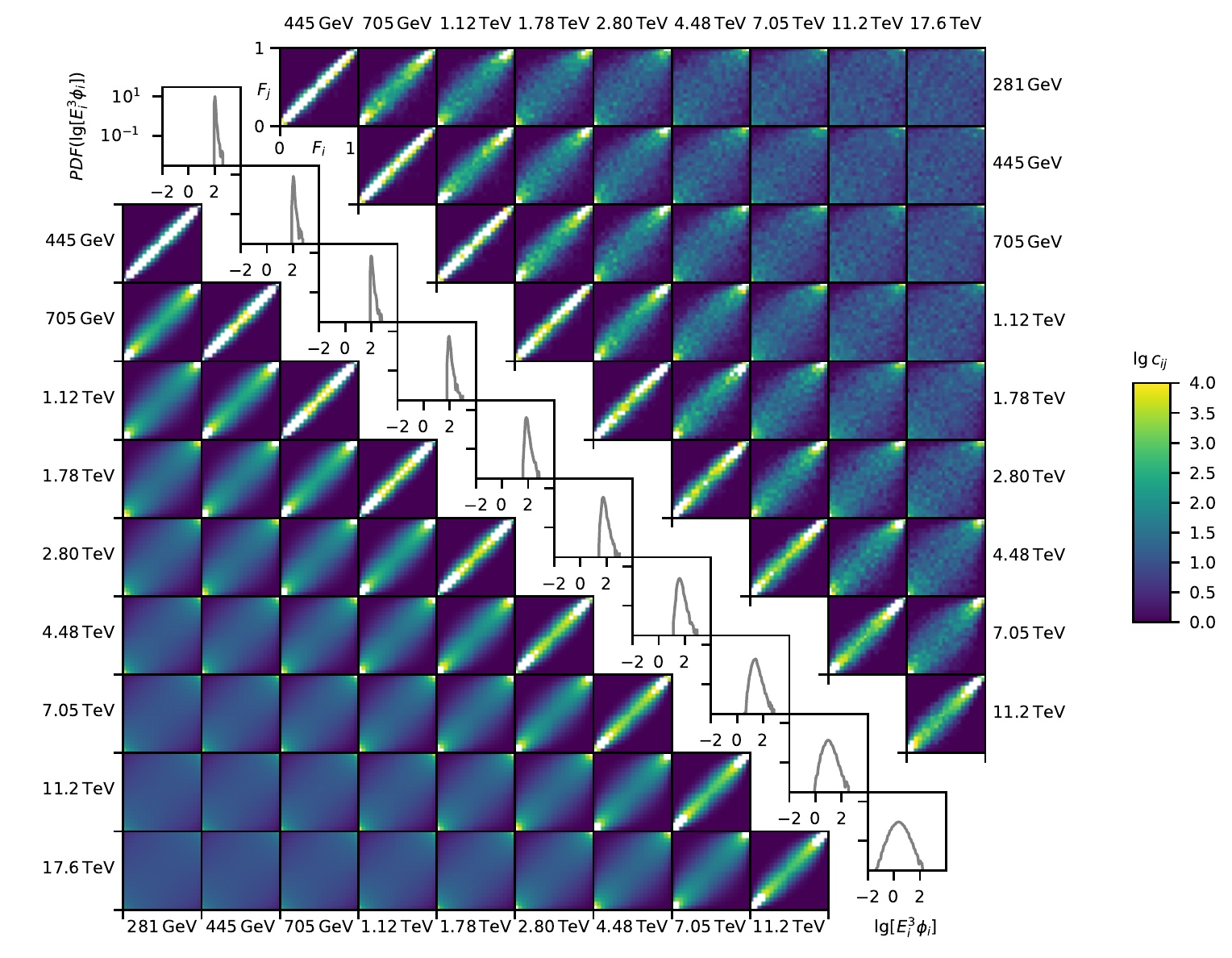}
\caption{Some pair copulae and marginals used in the pair-copula construction of the joint density, for a Galactic source rate of $2 \times 10^4 \, \text{Myr}^{-1}$. The upper triangle matrix shows 2D histograms of the transformed random variables $F_i$. Each panel corresponds to two energies, $E_i$ and $E_j$, as indicated on the top and right of the panels. The lower triangle matrix shows the copula density fitted to the MC sample, again with one panel for each combination of energies $E_i$ and $E_j$, indicated on the bottom and left of the panels. Finally, the diagonal shows the 1D marginals of the scaled flux logarithm, $\lg \left( E^3 \phi [\mathrm{GeV}^2 \, \mathrm{cm}^{-2} \, \mathrm{s}^{-1} \, \mathrm{sr}^{-1}] \right)$, each panel at a different energy $E_i$. Note that whereas in the copula plots the $x$- and $y$-axes are linear and go from $0$ to $1$, for the marginals, both axes are logarithmic.}
\label{fig:triangle_1e4}
\end{figure}

In Fig.~\ref{fig:triangle_1e4}, we show a graphical representation of some of the information contained in our copula based likelihood. The upper triangle matrix shows normalised 2D histograms for the transformed random variables $\hat{F}_i$ (cf.\ eq.~\ref{eqn:eCDF}), each row and column representing one particular energy bin. The histograms show the strong correlation between the fluxes in nearby energy bins (the histogram density is highest around the diagonal) whereas for energies far apart the fluxes are less correlated (the 2D histograms are almost completely flat). The lower triangle matrix shows the copula density from our pair-copula construction using Gaussian bivariate copulae as fit to the MC sample. It can be seen to nicely reproduce the correlation structure of the MC sample. Finally, on the diagonal, we show the marginal densities as determined from histograms of fluxes at individual energies. The marginal distributions confirm the trend we already read off Fig.~\ref{fig:spec_1e4}: At low energies, the distribution of the logarithm of fluxes is rather narrow, but asymmetric. At higher energies, it becomes more symmetric and exhibits the power law tails that are characteristic of stable distributions.

\begin{figure}
\centering
\includegraphics[width=0.8\textwidth]{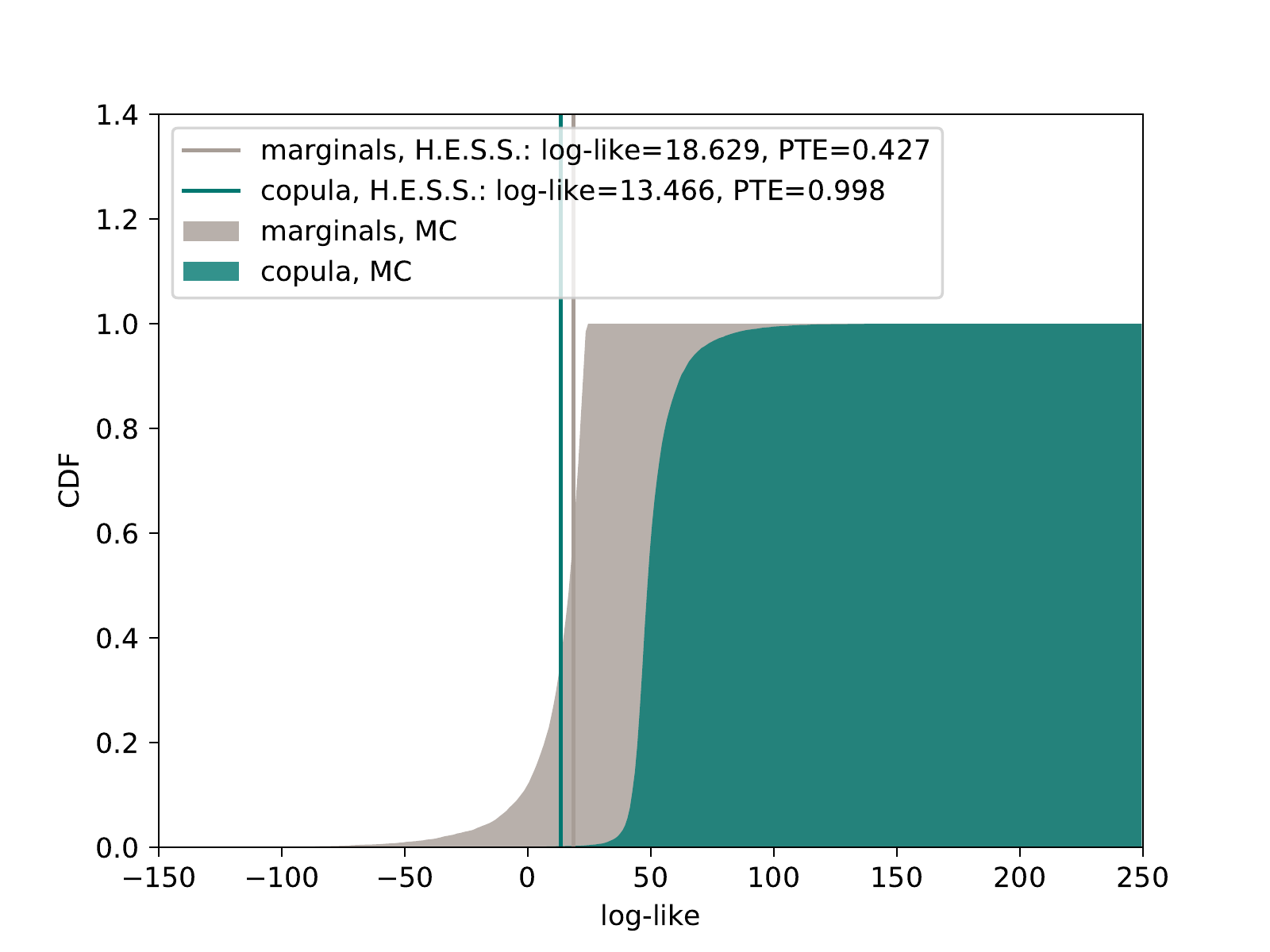}
\caption{Cumulative densities of the log-likelihoods in the MC computation for a Galactic source rate of $2 \times 10^4 \, \text{Myr}^{-1}$. The contributions from the marginal densities, $\log \mathcal{L}_{\text{marginals}}$, and from the copula, $\log \mathcal{L}_{\text{copula}}$, are shown in grey and dark cyan, respectively. The vertical grey and dark cyan lines indicate the marginal and copula log-likelihoods of the H.E.S.S. broken power law.}
\label{fig:hist_1e4}
\end{figure}

We have evaluated the likelihood for the H.E.S.S. broken power law spectrum by substituting $\vec{\hat{\phi}}$, cf.\ eq.~\ref{eqn:HESS_broken_power_law}, into eq.~\ref{eqn:marginals_copula} making use of eq.~\ref{eqn:likelihood}, and have found a contribution to the log-likelihood from the marginals of $\log \mathcal{L}_{\text{marginals}} \simeq 19$ and from the copula of $\log \mathcal{L}_{\text{copula}} \simeq 13$. The total log-likelihood is just the sum, $\log \mathcal{L}_{\text{total}} \simeq 32$. To evaluate the statistical compatibility between model and data, we have to compare theses numbers to the expected distribution of log-likelihoods in the statistical ensemble. We have therefore also substituted the spectra $\vec{\phi}_s$ for all members of the ensemble into the likelihood function, and we plot the CDFs of their contributions in Fig.~\ref{fig:hist_1e4}. It is noticeable that both distributions show deviations from the CDF of a $\chi^2$-distribution that we would have expected for a Gaussian joint density. In particular, the CDF of the $\log \mathcal{L}_{\text{marginals}}$ is very sharp around $\log \mathcal{L}_{\text{marginals}} \approx 20$, meaning that most realisations lead to spectra with log-likelihood $\lesssim 20$, but then there is a long tail of realisations with much smaller likelihood.

We are now ready to estimate how ``typical'' the H.E.S.S. broken power law spectrum is in a stochastic source model. For $\log \mathcal{L}_{\text{marginals}} \simeq 19$, we find a probability to exceed (PTE) of $0.43$, meaning that the stochastic model is giving a good fit to the broken power law within the statistical uncertainty induced by source discreteness, without overfitting. In contrast, for $\log \mathcal{L}_{\text{copula}} = 13$ we find a large PTE of $0.998$, meaning that the members of the random ensemble show too little variations from the median. As the distribution of log-likelihoods is dominated by the copula contribution, we conclude that the total log-likelihood is also incompatible with the distribution of total log-likelihoods.

We conclude that the stochastic source model with the source rate $\mathcal{R} = 2 \times 10^4 \, \text{Myr}^{-1}$ (and the other parameters shown in Tbl.~\ref{tbl:parameters}) is statistically disfavoured. Next, we will turn to a simple modification that makes the stochastic source model statistically perfectly acceptable.

\subsection{Reduced supernova rate}
\label{sec:reduced}

\begin{figure}[!tb]
\centering
\includegraphics[width=0.8\textwidth]{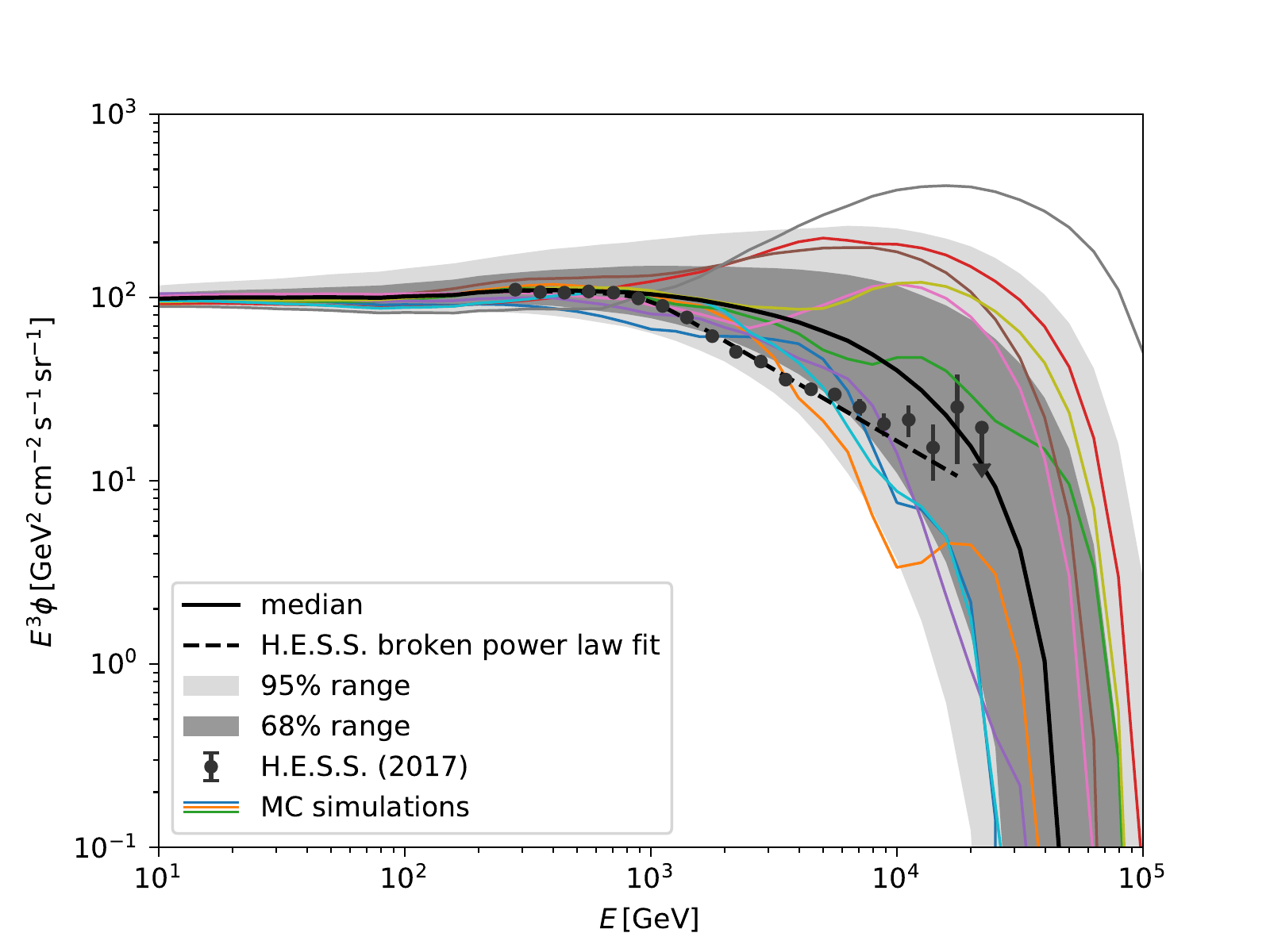}
\caption{Same as Fig.~\ref{fig:spec_1e4}, but for the reduced source rate, $\mathcal{R} = 10^3 \, \text{Myr}^{-1}$. (See Tbl.~\ref{tbl:parameters} for the other parameters used.)}
\label{fig:spec_1e3}
\end{figure}

\begin{figure}[!thb]
\centering
\includegraphics[scale=1]{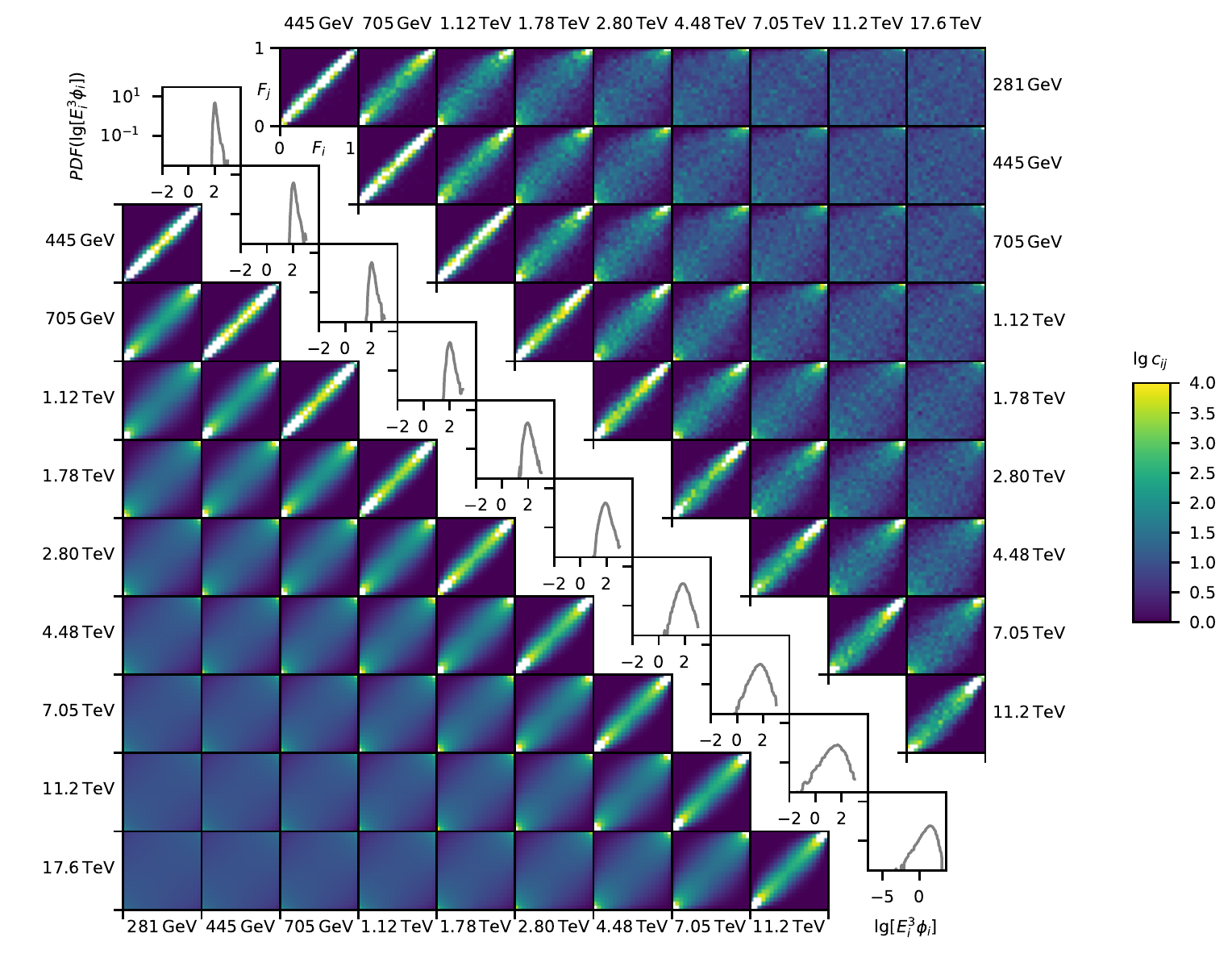}
\caption{Same as Fig.~\ref{fig:triangle_1e4}, but for the reduced source rate, $\mathcal{R} = 10^3 \, \text{Myr}^{-1}$. (See Tbl.~\ref{tbl:parameters} for the other parameters used.) Note the changing $x$-axis range in the 1D marginals at the highest energy.}
\label{fig:triangle_1e3}
\end{figure}

In the previous section, we found that the stochastic model with a source rate $\mathcal{R} = 2 \times 10^4 \, \text{Myr}^{-1}$ is statistically disfavoured, given the measurement of a broken power law by H.E.S.S. and confirmed directly by DAMPE. More specifically, the contribution to the likelihood from the copula is too small, indicating that there is too little variation between energy intervals. A simple way to fix both issues is to reduce the source rate: This will increase the variation in the spectrum between nearby energy bins, thus decreasing the likelihood contribution from the copula. We decided to reduce the source rate by one order of magnitude to $\mathcal{R} = 2 \times 10^3 \, \text{Myr}^{-1}$.

In Fig.~\ref{fig:spec_1e3}, we present the results from the MC sample with the reduced source rate in the same way as in Fig.~\ref{fig:spec_1e4}. Comparing with Fig.~\ref{fig:spec_1e4} it is evident that the level of variation between different energy bins and between different realisations of the flux is indeed significantly increased. As the mean turnover in energy is related to the mean distance to the nearest young source, we had to increase the source cut-off energy to $E_{\text{cut}} = 10^5 \, \text{GeV}$ in order for the roll-over not to occur at too low an energy.

Fig.~\ref{fig:triangle_1e3} shows the histograms of transformed random variables, $F_i$, 2D marginals of the copula and 1D marginals. Comparing with Fig.~\ref{fig:triangle_1e4}, which shows the same for the original source rate, confirms our previous impression: Both the 1D PDFs and the 2D copula densities are wider for the reduced source rate.

\begin{figure}[thb]
\centering
\includegraphics[width=0.8\textwidth]{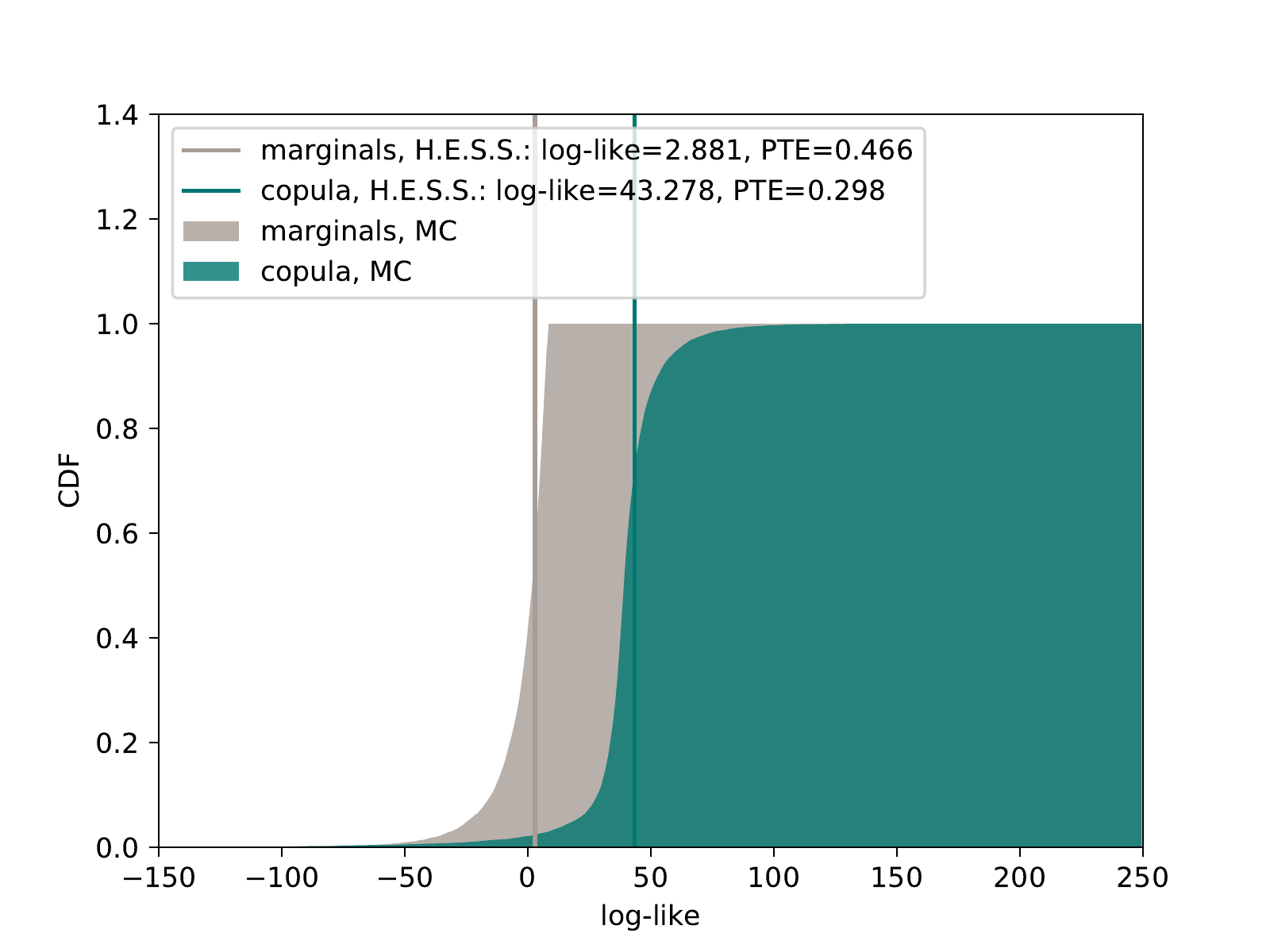}
\caption{Same as Fig.~\ref{fig:hist_1e4}, but for the reduced source rate, $\mathcal{R} = 10^3 \, \text{Myr}^{-1}$. (See Tbl.~\ref{tbl:parameters} for the other parameters used.)}
\label{fig:hist_1e3}
\end{figure}

Finally, we determine the marginal and copula contributions to log-likelihood for the H.E.S.S. broken power law fit, $\log \mathcal{L}_{\text{marginals}} \simeq 2.9$ and $\log \mathcal{L}_{\text{copula}} \simeq 43$, and compare it with the distribution of log-likelihoods in our MC sample, see Fig.~\ref{fig:hist_1e3}. Unlike for the case with the original source rate, cf.\ Fig.~\ref{fig:hist_1e4}, the PTEs are now perfectly mundane, $\text{PTE}_{\text{marginals}} \simeq 0.47$ and $\text{PTE}_{\text{copula}} \simeq 0.30$. The total log-likelihood, $\log \mathcal{L}_{\text{total}} \simeq 46$ is also compatible with the distribution of total log-likelihoods. We conclude that the stochastic source model with the reduced source rate of $\mathcal{R} = 2 \times 10^3 \, \text{Myr}^{-1}$ is statistically compatible with the H.E.S.S. broken power law.

\section{Discussion}
\label{sec:discussion}

\subsection{Justification of decreasing the source rate}
In the previous section, we have concluded that while a stochastic model with a source rate $\mathcal{R} = 2 \times 10^4 \, \text{Myr}^{-1}$ is statistically incompatible with the broken power law found by the H.E.S.S. collaboration, a model with a reduced rate of $\mathcal{R} = 2 \times 10^3 \, \text{Myr}^{-1}$ is statistically compatible. This modification of the source rate is rather extreme and certainly not justified by the uncertainty on the Galactic supernova rate. It can however be justified as an effective rate when considering the correlation in time and space of supernova events.

The source density as described in Sec.~\ref{sec:MC} has structure on spatial scales of kiloparsecs which are associated with the spiral structure of the Galaxy. However, there should also be structure on smaller scales due to the presence of massive gas clouds which are the host sites of star formation and supernova activity. The exact positions and temporal histories of these gas clouds cannot be known and must therefore also be treated as random variables. This reasoning results in a rather complicated, hierarchical random field for the source locations and times, with the position and times of gas clouds drawn from the large-scale gas density in the Galaxy in a first step, and with positions and times of individual sources drawn from the density profiles and histories of individual gas clouds in a second step.

An economical approximation of this hierarchical model would be to just increase the number of sources at each position and time drawn from the source density--or equivalently by reducing the source rate and increasing the source normalisation. While the positions and times will be wrong on the scales of individual gas clouds, this does not matter for the transport to the observer which is happening on much larger spatial and temporal scales. Decreasing the source rate $\mathcal{R}$ by one order of magnitude corresponds to an average number of 10 supernova explosion happening very nearby in space and time. (See also Ref.~\cite{Genolini:2016hte} for a discussion on this.)

\subsection{The issue with catalogues}

An approach oftentimes adopted in the literature is to use (all) available information on the distances to and ages of known (potential) sources of CR electrons in predicting the total observed flux via eq.~\ref{eqn:total}. For a given catalogue of sources, e.g.\ the ATNF catalogue for pulsars~\cite{Manchester:2004bp}, the model is deterministic. In that case, the goodness of fit can be quantified in the usual way: If the experimental errors are Gaussian, the residuals should be normal-distributed, and the $\chi^2$ test statistics should be $\chi^2$-distributed.

In practice, this approach can lead to a problem. Observations of sources in electromagnetic radiation, oftentimes in radio, suffer from incompleteness. In particular, the completeness of catalogues drops sharply beyond a certain distance and more importantly beyond a certain age. It was suggested a while ago~\cite{Kobayashi:2003kp} to adopt a somewhat hybrid approach: Adopt sources closer than a certain distance $s_1$ from catalogues and model sources further away than $s_1$ with a continuous source distribution. The problem with this kind of approach occurs due to incompleteness in the age distribution of sources:  We will certainly be missing sources due to their finite life times and the difference between propagation of electromagnetic radiation (ballistic) and CR electrons (diffusive). Even in the most favourable case, where particles are injected near the end of the source life time (e.g.\ for supernova remnants, shortly before the shell breaks up), and at high energies (where the diffusion time is shorter), we will not be able to observe sources older than, say, $10^5 \, \text{yr}$. Yet, TeV CR electrons can survive ($\tau_{\text{cool}} \sim 3 \times 10^5 \, \text{yr}$) and with a diffusion coefficient of $10^{29} \, \text{cm}^2 \text{s}^{-1}$ they still contribute significantly to the local flux. Ignoring the presence of these sources will therefore lead to an underestimated total flux, in particular at low energies.

\begin{figure}
\centering
\includegraphics[width=0.6\textwidth]{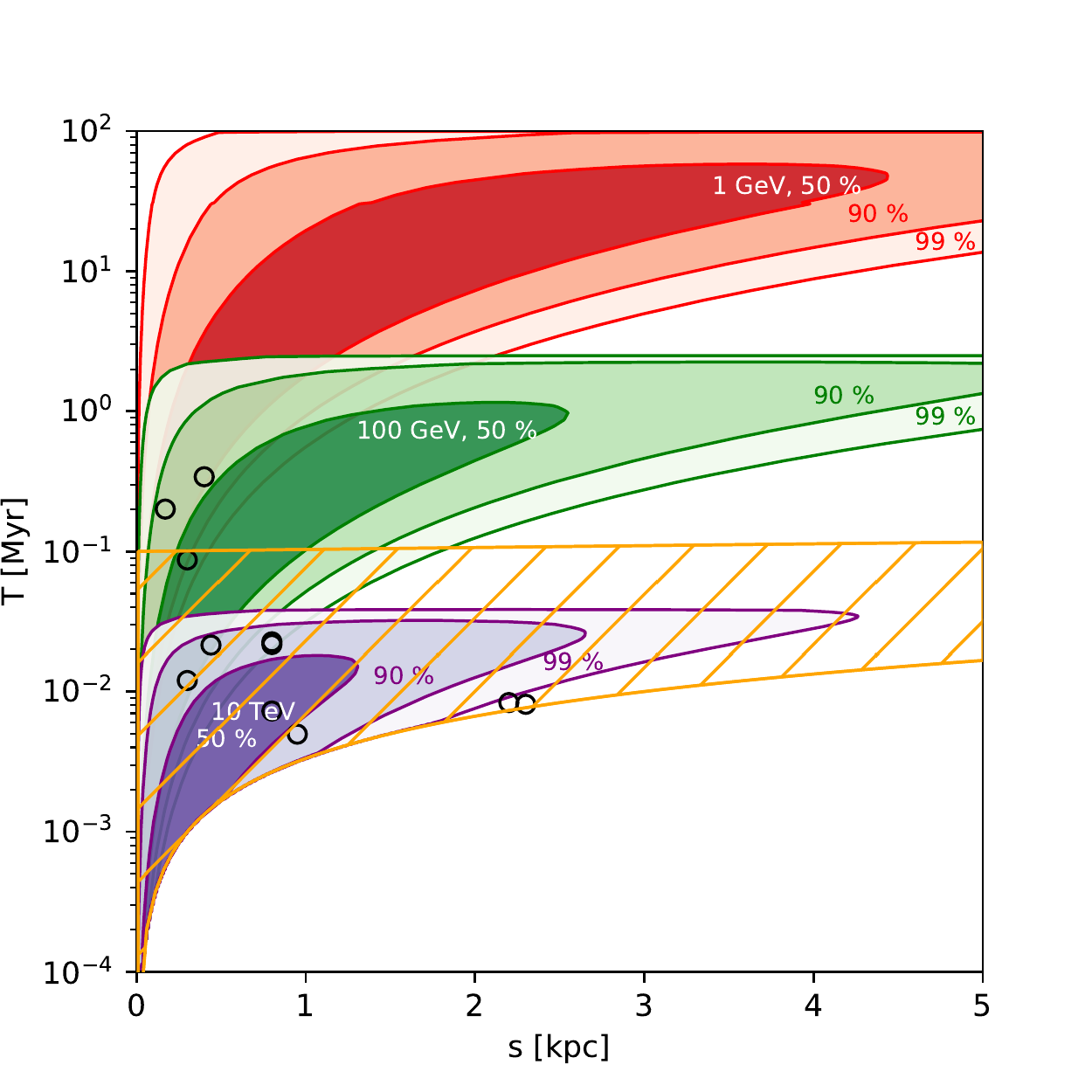}
\caption{Distance-age diagram and the contributions to electron fluxes at different energies. The shaded contours show the regions contributing $50 \, \%$, $90 \, \%$ and $99 \, \%$ of the fluxes at the energies of $E=1 \, \text{GeV}$, $100 \, \text{GeV}$ and $10 \, \text{TeV}$. (See text for how they have been generated.) The orange hatched band contains the sources which could be known from surveys in electromagnetic radiation. The open black circles indicate the subset of nearby sources from Green's catalogue of supernova remnants~\cite{Green:2015isa}.}
\label{fig:cum_density}
\end{figure}

We illustrate this problem with the help of Fig.~\ref{fig:cum_density}. For three different energies, $E=1 \, \text{GeV}$, $100 \, \text{GeV}$ and $10 \, \text{TeV}$, the shaded contours contain the regions that contribute $50 \, \%$, $90 \, \%$ and $99 \, \%$ of the flux at that particular energy. These contours have been generated in the following way: For a fixed energy $E$, we have multiplied the Green's function $\psi(s_i, T_i, E)$ with the source density $\sigma(s_i, T_i)$. The latter factorises into a constant source rate and, for simplicity, a homogeneous distribution in the two-dimensional disk for the spatial distribution. Starting from the maximum of this product density, we have drawn isodensity contours at $0.5$, $0.9$ and $0.99$ of the integrated density. The contours confirm our understanding from Sec.~\ref{sec:transport} about which distances and ages contribute to the electron flux at certain energies: At higher energies, sources must be necessarily younger, but are also closer on average in agreement with the diffusive relation $s^2 \sim \kappa T$. In Fig.~\ref{fig:cum_density}, we indicate the region that is populated by sources that we can know about in principle by the orange hatched region. Here, we consider $0.1 \, \text{Myr}$ as the maximum age characteristic for supernova remnants. Sources below this band are too young for their electromagnetic radiation to have reached us. Sources above (that is sources older than $0.1 \, \text{Myr}$), will have ceased to exist, thus not emitting electromagnetic radiation anymore, hence we cannot know about them. We have also indicated a subset of known nearby supernova remnants from Green's catalogue~\cite{Green:2015isa} by the open black circles. It is evident that sources that are too old for us to know about can still contribute to the electron flux at energies as high as $100 \, \text{GeV}$.

\begin{figure}
\centering
\includegraphics[width=0.7\textwidth]{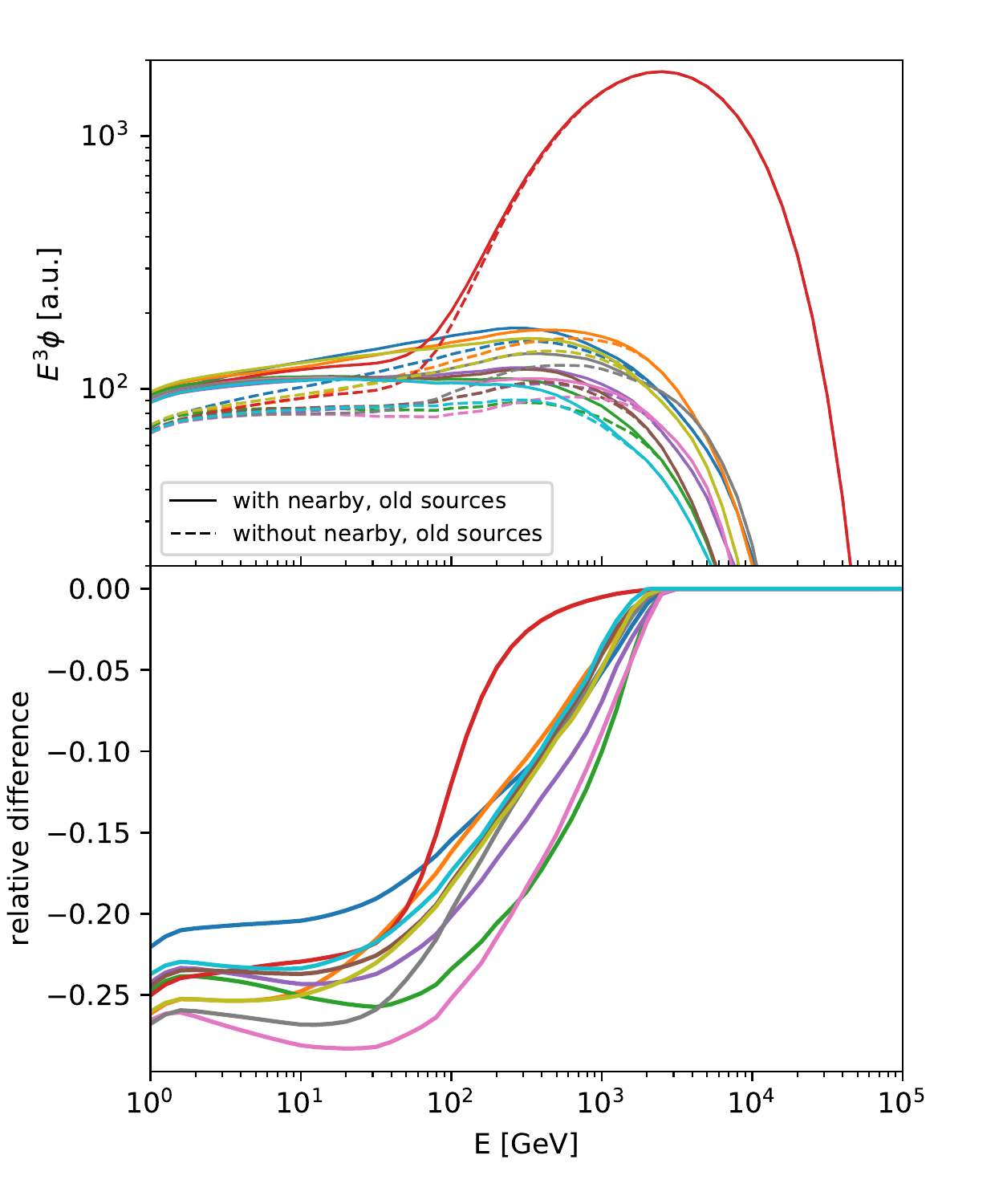}
\caption{\textbf{Top:} Total flux from our MC simulation, assuming a homogeneous source density in a disk around the observer and a constant source rate. Each of the ten solid, coloured lines corresponds to one random realisation of the source distribution. For the dashed lines, we have started from the same realisation, but removed those sources older than $0.1 \, \text{Myr}$, emulating the effect of catalogue incompleteness. \textbf{Bottom:} The relative difference between the random realisation without and with the sources older than $0.1 \, \text{Myr}$.}
\label{fig:problem_catalogues}
\end{figure}

To illustrate this problem, we return to the MC computation of electron fluxes that we used before to fit the copula parameters. Again, we assume an ensemble of galaxies, each with its own realisation of a distribution of sources. For each realisation, we compute the flux from each source and sum those fluxes up. In the upper panel of Fig.~\ref{fig:problem_catalogues}, the solid coloured lines show examples for the fluxes from 10 different realisations of the source distribution for source rate $\mathcal{R} = 2 \times 10^4 \, \text{Myr}^{-1}$, that is the canonical supernova rate in the Milky Way.

We now return to the problem at hand, that is the effect of source completeness in age. We adopt the above estimate of $t_{\text{max}} = 0.1 \, \text{Myr}$ for the age of the oldest sources, and remove from the ensemble all sources older than this. The resulting example spectra are shown by the dashed coloured lines in the upper panel of Fig.~\ref{fig:problem_catalogues}. In the lower panel we show the relative difference of fluxes from the cut source distribution and the full distribution of sources. We note that up to a few hundred GeV, cutting out the sources older than $t_{\text{max}}$ leads to a flux underestimated by tens of percent. This is to be compared to the experimental errors of a few percent at most from direct observations~\cite{Aguilar:2014fea,Aguilar:2014mma}. We conclude that the true model would lead to a bad fit or that if we allowed the model to adjust its parameters and it would lead to a good fit, it would not be the correct model.

There are a number of other potential issues when using catalogues for predicting the spectrum. Oftentimes the uncertainties of the distances and ages are significant, but are not properly propagated into uncertainties in the predicted fluxes. Moreover, in the anisotropic, one-dimensional diffusion case, the straight line distance to a source can differ from the magnetic distance that is relevant for the diffusion of charged particles. Even worse, depending on magnetic topology, the observer might only be connected to a limited number of local sources. All this would break the connection between the (distance in the) catalogue and (those of) the set of relevant sources.

We conclude that the use of catalogues for predictions of the locally observed fluxes is unreliable and we are led to abandon the deterministic model based on the perfect knowledge of source distances and ages.

\subsection{Sensitivity to propagation parameters}

The results presented in Sec.~\ref{sec:results} rely on a MC-based analysis of the H.E.S.S. data. As such, the dependence of the ensemble of electron fluxes on transport and source spectral parameters are implicit and cannot simply be read off a formula. While the parameters that we adopted for the sources (i.e.\ $Q_*$, $\Gamma$ and $E_{\text{cut}}$) and for the transport of CR electrons ($z_{\text{max}}$, $\kappa_*$ and $\delta$) give acceptable marginal and copula log-likelihoods for the reduced source rate (see Sec.~\ref{sec:reduced}), we cannot exclude at this point that a different combination of parameters would also give acceptable log-likelihoods. Short of running a full parameter study, we provide a few comments on the sensitivity of our conclusions to variations of the transport and source spectral parameters.

\begin{itemize}
\item Increasing (decreasing) $\kappa_*$ increases (decreases) the diffusion loss-length $\ell^2$, therefore more (fewer) sources contribute to the local fluxes. Of course, their normalisation can be adjusted by decreasing (increasing) $Q_*$, but the fluctuations will be smaller (larger), both at individual energies and in the correlations.
\item While the spectra at high energies become insensitive to the half-height of the CR halo, $z_{\text{max}}$, we note that in order to match nuclear secondary-to-primary ratios like the Boron-to-Carbon ratio, requires adjusting $\kappa_*$, due to the well-known degeneracy between $\kappa_*$ and $z_{\text{max}}$.
\item Another obvious degeneracy is the one between the source spectral index $\Gamma$ and the spectral index of the diffusion coefficient, $\delta$. In order to match the observed flux $\propto E^{-3}$ below $\sim 1 \, \text{TeV}$ requires that one be increased when the other is decreased and vice versa. Of course, this is only possible within certain limits set by other means, e.g.\ bounds on $\delta$ from the Boron-to-Carbon ratio. Note, however, that the fluctuations have a stronger dependence on $\delta$ then the median of fluxes for the same reason that the level of fluctuations depends on $\kappa_*$.
\item Finally, we expect that varying $E_{\text{cut}}$ affects the flux median and the fluctuations in about the same way. \end{itemize}
Note that some of these dependencies can be gleaned by inspection of the analytical approximations by equations from Ref.~\cite{Mertsch:2010fn}, in particular their eq.~3.16 for the flux expectation value and their eq.~4.13 that encodes the size of the fluctuations at individual energies.

As already discussed in Sec.~\ref{sec:StatisticalModel}, there is a certain degeneracy between the transport and source spectral parameters on the one hand and the particular realisation of the source distribution on the other hand, in the sense that the flux from one combination of those (e.g.\ with soft spectra and one particular realisation of the source distribution) can approximate the flux from a different combination (e.g.\ with hard spectra and a different realisation of the source distribution). What our approach can provide is a statistical verdict on how likely a particular realisation is, based on the distribution of fluxes at individual energies and their correlations.

\section{Summary and conclusion}
\label{sec:SummaryConclusion}

We have presented a stochastic model for the local all-electron spectrum that is in agreement with the broken power law recently observed by H.E.S.S. The stochastic nature of the spectrum is due to the assumed random nature of discrete sources of CR electrons. We have run MC simulations, adding up the contributions from $10^6$ sources and scanning over $10^4$ random realisations of different source distributions. We have adopted source rates of $\mathcal{R} = 2 \times 10^4 \, \text{Myr}^{-1}$ and $\mathcal{R} = 2 \times 10^3 \, \text{Myr}^{-1}$, the former in agreement with the Galactic supernova rate, the latter one order of magnitude lower which served as a simple proxy for sources that are spatially and temporally correlated. For the first time, we have not only quantified the distribution of the local flux at individual energies, but also considered the correlations between different energy bins. To this end, we have parametrised the joint density of the fluxes in different energy bins by the product of 1D marginal distributions and a copula. We have employed a simplified vine copula model that factorises the copula into products of bivariate copulas. The free parameters of the pair copulae were inferred by fitting the copula to the sample of transformed random variables from the MC simulation. We have found that using Gaussian pair copulae gives a better description of the MC sample than using Clayton survival copulae or Gumbel copulae. Eventually, we have computed the log-likelihood of the broken power law fit presented by the H.E.S.S. collaboration and compared it to the range of log-likelihoods expected from the MC sample. We have found the model with $\mathcal{R} = 2 \times 10^4 \, \text{Myr}^{-1}$ to exhibit too little variation between energy bins whereas the model with $\mathcal{R} = 2 \times 10^3 \, \text{Myr}^{-1}$ is perfectly compatible with the H.E.S.S. data. Finally, we have critically reviewed the alternative approach of using catalogues for predicting the all-electron spectrum, concluding that due to issues with catalogue completeness the predicted spectra become unreliable below energies of a few hundred GeV.

\appendix
\section{Pair copulae}
\label{sec:appendix}
In the following, we list the copula densities and $h$-functions of these three copula families for reference~\cite{Venter:2001,Aas:2009} and show contour plots. See Ref.~\cite{Schirmacher:2008} for a gallery of scatter plots of other popular pair copula types.

\subsection{Gaussian copula}

\begin{figure}[b]
\includegraphics[scale=1]{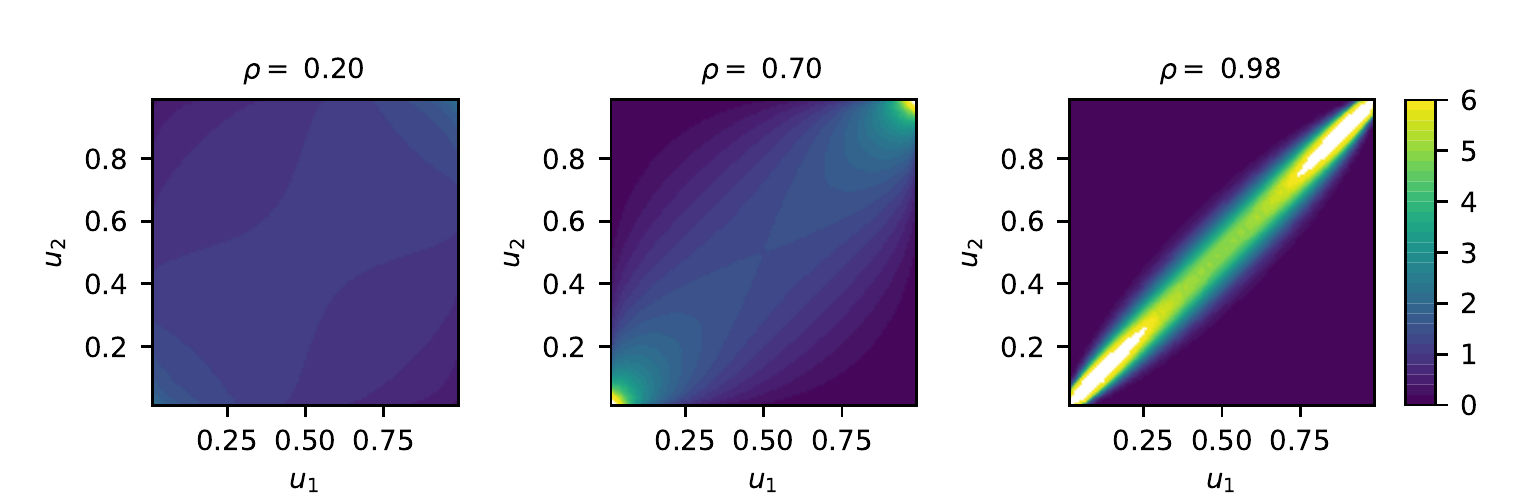}
\caption{Contour plots of the bivariate Gaussian copula with parameters $\rho=0.2$ (left), $0.7$ (middle) and $0.98$ (right).}
\label{fig:paircoupla_Gaussian}
\end{figure}

The Gaussian copula has the density
\begin{equation}
c(u_1, u_2) = \frac{1}{\sqrt{1 - \rho^2}} \exp \left[ - \frac{\rho^2 (x_1^2 + x_2^2) - 2 \rho x_1 x_2}{2 (1 - \rho^2)} \right] \, ,
\label{eqn:Gaussian_copula_density}
\end{equation}
and the $h$-function is
\begin{equation}
h(u_1, u_2) = \Phi \left( \frac{x_1 - \rho x_2}{\sqrt{1 - \rho^2}} \right) \, .
\end{equation}
Here, $x_i$ = $\Phi^{-1}(u_i)$, and $\Phi( \cdot )$ and $\Phi^{-1}( \cdot )$ are the standard normal CDF and its inverse, respectively. Fig.~\ref{fig:paircoupla_Gaussian} shows contour plots of the bivariate Gaussian copula with various choices for the parameter $\rho$.

\subsection{Gumbel copula}

\begin{figure}
\includegraphics[scale=1]{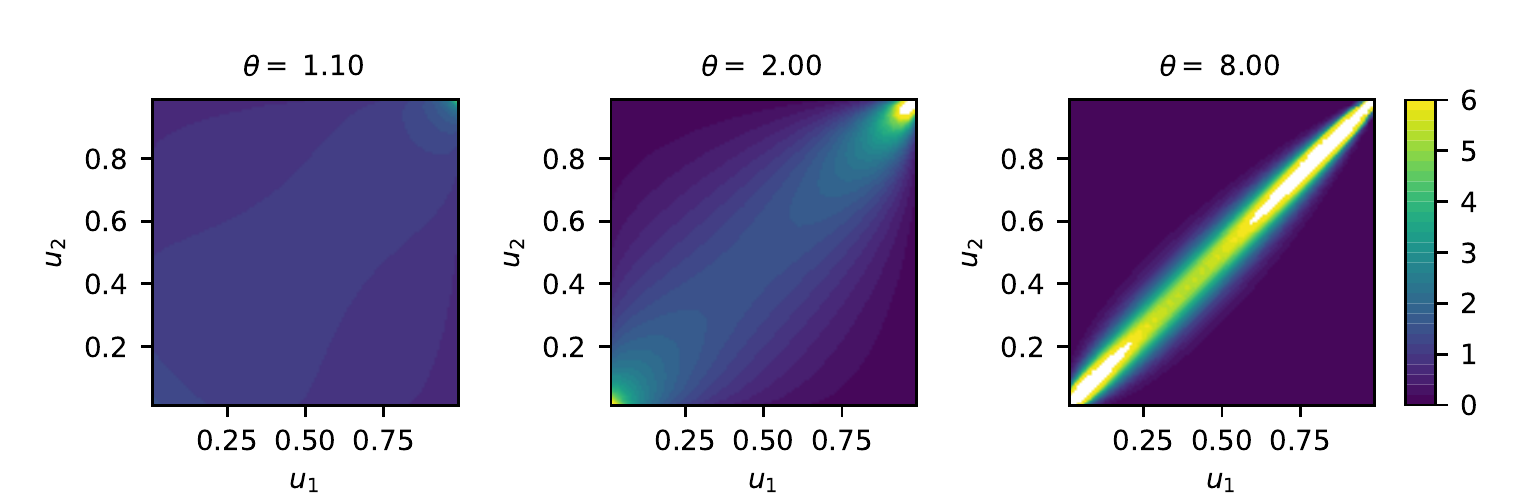}
\caption{Contour plots of the bivariate Gumbel copula with parameters $\theta=1.1$ (left), $2$ (middle) and $8$ (right).}
\label{fig:paircoupla_Gumbel}
\end{figure}

The Gumbel copula has the density,
\begin{align}
c(u_1, u_2) =& \exp \left[ - \left( (- \log u_1)^{\theta} + (- \log u_2)^{\theta} \right)^{1/\theta} \right] (u_1 u_2)^{-1} \left( (- \log u_1)^{\theta} + (- \log u_2)^{\theta} \right)^{-2+2/\theta} \nonumber 
\\ & \times \left( \log u_1 \log u_2 \right)^{\theta - 1} \left( 1 + (\theta - 1) \left( (- \log u_1)^{\theta} + (- \log u_2)^{\theta} \right)^{-1/\theta} \right) \, ,
\label{eqn:Gumbel_copula_density}
\end{align}
and the $h$-function is
\begin{align}
h(u_1, u_2) =& \exp \left[ - \left( (- \log u_1)^{\theta} + (- \log u_2)^{\theta} \right)^{1/\theta} \right] \frac{1}{u_2} (- \log u_2)^{\theta-1} \nonumber 
\\ & \times \left( (- \log u_1)^{\theta} + (- \log u_2)^{\theta} \right)^{1/\theta - 1} \, .
\end{align}
Fig.~\ref{fig:paircoupla_Gumbel} shows contour plots of the bivariate Gumbel copula with various choices for the parameter $\theta$.

\subsection{Clayton survival copula}

\begin{figure}
\includegraphics[scale=1]{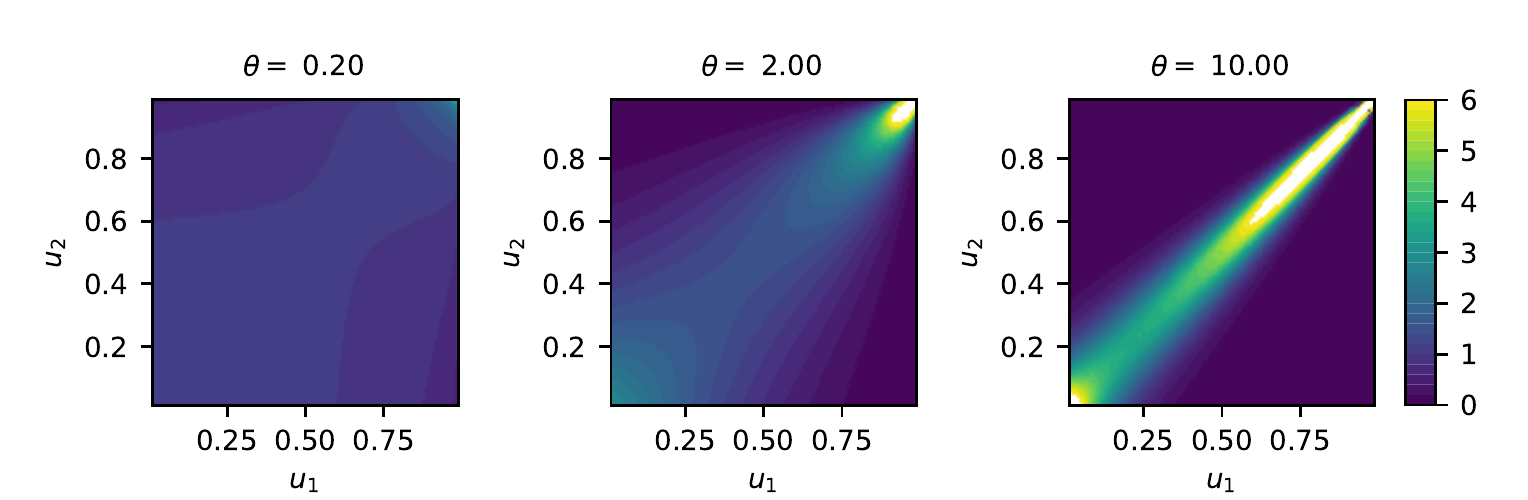}
\caption{Contour plots of the bivariate Clayton survival copula with parameters $\theta=0.2$ (left), $2$ (middle) and $10$ (right).}
\label{fig:paircoupla_Clayton_survival}
\end{figure}

The Clayton survival copula can be obtained from the usual Clayton copula by the substitution $u_i \to (1-u_i)$. Its density is
\begin{equation}
c(u_1, u_2) = (1 + \theta) \left( (1-u_1)(1-u_2) \right)^{-1-\theta} \left((1-u_1)^{-\theta} + (1-u_2)^{-\theta} - 1 \right)^{-1/\theta - 2} \, ,
\end{equation}
and the $h$-function is
\begin{equation}
h(u_1, u_2) = 1 - (1-u_2)^{-1-\theta} \left((1-u_1)^{-\theta} + (1-u_2)^{-\theta} - 1 \right)^{-1/\theta - 1} \, .
\label{eqn:Clayton_survival_copula_h}
\end{equation}
Fig.~\ref{fig:paircoupla_Clayton_survival} shows contour plots of the bivariate Clayton survival copula with various choices for the parameter $\theta$.

\bibliographystyle{jhep}
\bibliography{correlations}

\providecommand{\href}[2]{#2}\begingroup\raggedright\begin{thebibliography}{10}

\bibitem{Hess:1912srp}
V.~F. Hess, \emph{{Über Beobachtungen der durchdringenden Strahlung bei sieben
  Freiballonfahrten}}, {\emph{Phys. Z.} {\bfseries 13} (1912) 1084}.

\bibitem{1977SPhD...22..327K}
G.~F. {Krymskii}, \emph{{A regular mechanism for the acceleration of charged
  particles on the front of a shock wave}}, {\emph{Soviet Physics Doklady}
  {\bfseries 22} (1977) 327}.

\bibitem{1977ICRC...11..132A}
W.~I. {Axford}, E.~{Leer} and G.~{Skadron}, \emph{{The acceleration of cosmic
  rays by shock waves}},  in \emph{{Proceedings of the 15th International
  Cosmic Ray Conference, Plovdiv, Bulgaria}}, pp.~132--137, 1977.

\bibitem{Bell1978a}
A.~R. {Bell}, \emph{{The acceleration of cosmic rays in shock fronts. I}},
  \href{https://doi.org/10.1093/mnras/182.2.147}{\emph{Mon.~Not.~Roy.~Astron.~Soc.}
  {\bfseries 182} (1978) 147}.

\bibitem{Bell1978b}
A.~R. {Bell}, \emph{{The acceleration of cosmic rays in shock fronts. II}},
  \href{https://doi.org/10.1093/mnras/182.3.443}{\emph{Mon.~Not.~Roy.~Astron.~Soc.}
  {\bfseries 182} (1978) 443}.

\bibitem{Blandford1978}
R.~D. {Blandford} and J.~P. {Ostriker}, \emph{{Particle acceleration by
  astrophysical shocks}},
  \href{https://doi.org/10.1086/182658}{\emph{Astrophys.~J.~Lett.} {\bfseries
  221} (1978) L29}.

\bibitem{TheOriginofCosmicRays1964}
V.~L. {Ginzburg} and S.~I. {Syrovatskii}, \emph{{The Origin of Cosmic Rays}}.
  Macmillan, New York, 1964.

\bibitem{Funk:2015ena}
S.~Funk, \emph{{Ground- and Space-Based Gamma-Ray Astronomy}},
  \href{https://doi.org/10.1146/annurev-nucl-102014-022036}{\emph{Ann. Rev.
  Nucl. Part. Sci.} {\bfseries 65} (2015) 245}
  [\href{https://arxiv.org/abs/1508.05190}{{\ttfamily 1508.05190}}].

\bibitem{Genolini:2016hte}
Y.~Genolini, P.~Salati, P.~Serpico and R.~Taillet, \emph{{Stable laws and
  cosmic ray physics}},
  \href{https://doi.org/10.1051/0004-6361/201629903}{\emph{Astron. Astrophys.}
  {\bfseries 600} (2017) A68}
  [\href{https://arxiv.org/abs/1610.02010}{{\ttfamily 1610.02010}}].

\bibitem{2007BRASP..71..494P}
A.~D. {Panov}, J.~H. {Adams}, Jr., H.~S. {Ahn}, K.~E. {Batkov}, G.~L.
  {Bashindzhagyan}, J.~W. {Watts} et~al., \emph{{Elemental energy spectra of
  cosmic rays from the data of the ATIC-2 experiment}},
  \href{https://doi.org/10.3103/S1062873807040168}{\emph{Bulletin of the
  Russian Academy of Sciences, Physics} {\bfseries 71} (2007) 494}.

\bibitem{Ahn:2010gv}
H.~S. Ahn et~al., \emph{{Discrepant hardening observed in cosmic-ray elemental
  spectra}},
  \href{https://doi.org/10.1088/2041-8205/714/1/L89}{\emph{Astrophys. J.}
  {\bfseries 714} (2010) L89}
  [\href{https://arxiv.org/abs/1004.1123}{{\ttfamily 1004.1123}}].

\bibitem{Adriani:2011cu}
{\scshape PAMELA} collaboration, O.~Adriani et~al., \emph{{PAMELA Measurements
  of Cosmic-ray Proton and Helium Spectra}},
  \href{https://doi.org/10.1126/science.1199172}{\emph{Science} {\bfseries 332}
  (2011) 69} [\href{https://arxiv.org/abs/1103.4055}{{\ttfamily 1103.4055}}].

\bibitem{Aguilar:2015ooa}
{\scshape AMS} collaboration, M.~Aguilar et~al., \emph{{Precision Measurement
  of the Proton Flux in Primary Cosmic Rays from Rigidity 1 GV to 1.8 TV with
  the Alpha Magnetic Spectrometer on the International Space Station}},
  \href{https://doi.org/10.1103/PhysRevLett.114.171103}{\emph{Phys. Rev. Lett.}
  {\bfseries 114} (2015) 171103}.

\bibitem{Aguilar:2015ctt}
{\scshape AMS} collaboration, M.~Aguilar et~al., \emph{{Precision Measurement
  of the Helium Flux in Primary Cosmic Rays of Rigidities 1.9 GV to 3 TV with
  the Alpha Magnetic Spectrometer on the International Space Station}},
  \href{https://doi.org/10.1103/PhysRevLett.115.211101}{\emph{Phys. Rev. Lett.}
  {\bfseries 115} (2015) 211101}.

\bibitem{Aguilar:2017hno}
{\scshape AMS} collaboration, M.~Aguilar et~al., \emph{{Observation of the
  Identical Rigidity Dependence of He, C, and O Cosmic Rays at High Rigidities
  by the Alpha Magnetic Spectrometer on the International Space Station}},
  \href{https://doi.org/10.1103/PhysRevLett.119.251101}{\emph{Phys. Rev. Lett.}
  {\bfseries 119} (2017) 251101}.

\bibitem{Kumar:2014dma}
R.~Kumar and D.~Eichler, \emph{{Large-scale Anisotropy of TeV-band Cosmic
  Rays}}, \href{https://doi.org/10.1088/0004-637X/785/2/129}{\emph{Astrophys.
  J.} {\bfseries 785} (2014) 129}.

\bibitem{Mertsch:2014cua}
P.~Mertsch and S.~Funk, \emph{{Solution to the cosmic ray anisotropy problem}},
  \href{https://doi.org/10.1103/PhysRevLett.114.021101}{\emph{Phys. Rev. Lett.}
  {\bfseries 114} (2015) 021101}
  [\href{https://arxiv.org/abs/1408.3630}{{\ttfamily 1408.3630}}].

\bibitem{Kounine:2012ega}
A.~Kounine, \emph{{The Alpha Magnetic Spectrometer on the International Space
  Station}}, \href{https://doi.org/10.1142/S0218301312300056}{\emph{Int. J.
  Mod. Phys.} {\bfseries E21} (2012) 1230005}.

\bibitem{Rosier-Lees:2014}
S.~Rosier-Lees. {Proceedings of Astroparticle Physics TEVPA/IDM, Amsterdam},
  2014.

\bibitem{Ting:2013sea}
S.~Ting, \emph{{The Alpha Magnetic Spectrometer on the International Space
  Station}},
  \href{https://doi.org/10.1016/j.nuclphysbps.2013.09.028}{\emph{Nucl. Phys.
  Proc. Suppl.} {\bfseries 243-244} (2013) 12}.

\bibitem{Lee:2012}
S.-C. Lee. {Proceedings of the 20th International Conference on Supersymmetry
  and Unification of Fundamental Interactions (SUSY 2012), Beijing}, 2012.

\bibitem{Aguilar:2012}
M.~Aguilar. {Proceedings of the XL International Meeting on Fundamental
  Physics, Centro de Ciencias de Benasque Pedro Pascual}, 2012.

\bibitem{Schael:2012}
S.~Schael. {Proceedings of the 10th Symposium on Sources and Detection of Dark
  Matter and Dark Energy in the Universe, Los Angeles}, 2012.

\bibitem{Bertucci:2011}
B.~Bertucci Proc. Sci., EPS-HEP (2011) 67, 2011,
  \href{https://doi.org/10.22323/1.134.0067}{DOI}.

\bibitem{Incagli:2010zz}
{\scshape AMS} collaboration, M.~Incagli, \emph{{Astroparticle physiscs with
  AMS02}}, \href{https://doi.org/10.1063/1.3395995}{\emph{AIP Conf. Proc.}
  {\bfseries 1223} (2010) 43}.

\bibitem{Battiston:2008zza}
{\scshape AMS 02} collaboration, R.~Battiston, \emph{{The antimatter
  spectrometer (AMS-02): A particle physics detector in space}},
  \href{https://doi.org/10.1016/j.nima.2008.01.044}{\emph{Nucl. Instrum. Meth.}
  {\bfseries A588} (2008) 227}.

\bibitem{Torii:2015lck}
{\scshape CALET} collaboration, S.~Torii, \emph{{The CALorimetric Electron
  Telescope (CALET): a High-Energy Astroparticle Physics Observatory on the
  International Space Stati}},
  \href{https://doi.org/10.22323/1.236.0581}{\emph{PoS} {\bfseries ICRC2015}
  (2016) 581}.

\bibitem{TheDAMPE:2017dtc}
{\scshape DAMPE} collaboration, J.~Chang et~al., \emph{{The DArk Matter
  Particle Explorer mission}},
  \href{https://doi.org/10.1016/j.astropartphys.2017.08.005}{\emph{Astropart.
  Phys.} {\bfseries 95} (2017) 6}
  [\href{https://arxiv.org/abs/1706.08453}{{\ttfamily 1706.08453}}].

\bibitem{Abdollahi:2017nat}
{\scshape Fermi-LAT} collaboration, S.~Abdollahi et~al., \emph{{Cosmic-ray
  electron-positron spectrum from 7 GeV to 2 TeV with the Fermi Large Area
  Telescope}}, \href{https://doi.org/10.1103/PhysRevD.95.082007}{\emph{Phys.
  Rev.} {\bfseries D95} (2017) 082007}
  [\href{https://arxiv.org/abs/1704.07195}{{\ttfamily 1704.07195}}].

\bibitem{Malyshev:2009tw}
D.~Malyshev, I.~Cholis and J.~Gelfand, \emph{{Pulsars versus Dark Matter
  Interpretation of ATIC/PAMELA}},
  \href{https://doi.org/10.1103/PhysRevD.80.063005}{\emph{Phys. Rev.}
  {\bfseries D80} (2009) 063005}
  [\href{https://arxiv.org/abs/0903.1310}{{\ttfamily 0903.1310}}].

\bibitem{Cholis:2017ccs}
I.~Cholis, T.~Karwal and M.~Kamionkowski, \emph{{Features in the Spectrum of
  Cosmic-Ray Positrons from Pulsars}},
  \href{https://doi.org/10.1103/PhysRevD.97.123011}{\emph{Phys. Rev.}
  {\bfseries D97} (2018) 123011}
  [\href{https://arxiv.org/abs/1712.00011}{{\ttfamily 1712.00011}}].

\bibitem{Adriani:2008zr}
{\scshape PAMELA} collaboration, O.~Adriani et~al., \emph{{An anomalous
  positron abundance in cosmic rays with energies 1.5-100 GeV}},
  \href{https://doi.org/10.1038/nature07942}{\emph{Nature} {\bfseries 458}
  (2009) 607} [\href{https://arxiv.org/abs/0810.4995}{{\ttfamily 0810.4995}}].

\bibitem{Adriani:2011xv}
{\scshape PAMELA} collaboration, O.~Adriani et~al., \emph{{The cosmic-ray
  electron flux measured by the PAMELA experiment between 1 and 625 GeV}},
  \href{https://doi.org/10.1103/PhysRevLett.106.201101}{\emph{Phys. Rev. Lett.}
  {\bfseries 106} (2011) 201101}
  [\href{https://arxiv.org/abs/1103.2880}{{\ttfamily 1103.2880}}].

\bibitem{Adriani:2013uda}
{\scshape PAMELA} collaboration, O.~Adriani et~al., \emph{{Cosmic-Ray Positron
  Energy Spectrum Measured by PAMELA}},
  \href{https://doi.org/10.1103/PhysRevLett.111.081102}{\emph{Phys. Rev. Lett.}
  {\bfseries 111} (2013) 081102}
  [\href{https://arxiv.org/abs/1308.0133}{{\ttfamily 1308.0133}}].

\bibitem{Aguilar:2014fea}
{\scshape AMS} collaboration, M.~Aguilar et~al., \emph{{Precision Measurement
  of the ($e^+ + e^-$) Flux in Primary Cosmic Rays from 0.5 GeV to 1 TeV with
  the Alpha Magnetic Spectrometer on the International Space Station}},
  \href{https://doi.org/10.1103/PhysRevLett.113.221102}{\emph{Phys. Rev. Lett.}
  {\bfseries 113} (2014) 221102}.

\bibitem{Aguilar:2014mma}
{\scshape AMS} collaboration, M.~Aguilar et~al., \emph{{Electron and Positron
  Fluxes in Primary Cosmic Rays Measured with the Alpha Magnetic Spectrometer
  on the International Space Station}},
  \href{https://doi.org/10.1103/PhysRevLett.113.121102}{\emph{Phys. Rev. Lett.}
  {\bfseries 113} (2014) 121102}.

\bibitem{Aguilar:2018ons}
{\scshape AMS} collaboration, M.~Aguilar et~al., \emph{{Observation of Complex
  Time Structures in the Cosmic-Ray Electron and Positron Fluxes with the Alpha
  Magnetic Spectrometer on the International Space Station}},
  \href{https://doi.org/10.1103/PhysRevLett.121.051102}{\emph{Phys. Rev. Lett.}
  {\bfseries 121} (2018) 051102}.

\bibitem{Adriani:2017efm}
{\scshape CALET} collaboration, O.~Adriani et~al., \emph{{Energy Spectrum of
  Cosmic-Ray Electron and Positron from 10 GeV to 3 TeV Observed with the
  Calorimetric Electron Telescope on the International Space Station}},
  \href{https://doi.org/10.1103/PhysRevLett.119.181101}{\emph{Phys. Rev. Lett.}
  {\bfseries 119} (2017) 181101}
  [\href{https://arxiv.org/abs/1712.01711}{{\ttfamily 1712.01711}}].

\bibitem{Ambrosi:2017wek}
{\scshape DAMPE} collaboration, G.~Ambrosi et~al., \emph{{Direct detection of a
  break in the teraelectronvolt cosmic-ray spectrum of electrons and
  positrons}}, \href{https://doi.org/10.1038/nature24475}{\emph{Nature}
  {\bfseries 552} (2017) 63}
  [\href{https://arxiv.org/abs/1711.10981}{{\ttfamily 1711.10981}}].

\bibitem{WengICHEP2018}
{Z.~Weng [for the AMS Collaboration]}, \emph{{Precision Measurement of Electron
  and Positron Fluxes in Primary Cosmic Rays with the Alpha Magnetic
  Spectrometer on the International Space Station}},  Presentation at the 39th
  International Conference on High Energy Physics, Seoul, Korea, 2018,
  \href{https://indico.cern.ch/event/686555/contributions/2956951/}{https://indico.cern.ch/event/686555/contributions/2956951/}.

\bibitem{Aharonian:2008aa}
{\scshape H.E.S.S.} collaboration, F.~Aharonian et~al., \emph{{The energy
  spectrum of cosmic-ray electrons at TeV energies}},
  \href{https://doi.org/10.1103/PhysRevLett.101.261104}{\emph{Phys. Rev. Lett.}
  {\bfseries 101} (2008) 261104}
  [\href{https://arxiv.org/abs/0811.3894}{{\ttfamily 0811.3894}}].

\bibitem{Aharonian:2009ah}
{\scshape H.E.S.S.} collaboration, F.~Aharonian et~al., \emph{{Probing the ATIC
  peak in the cosmic-ray electron spectrum with H.E.S.S}},
  \href{https://doi.org/10.1051/0004-6361/200913323}{\emph{Astron. Astrophys.}
  {\bfseries 508} (2009) 561}
  [\href{https://arxiv.org/abs/0905.0105}{{\ttfamily 0905.0105}}].

\bibitem{KerszbergICRC2017}
{D.~Kerszberg, M.~Kraus, D.~Kolitzus, K.~Egberts, S.~Funk, J.-P.~Lenain,
  O.~Reimer, P.~Vincent [for the H.E.S.S. Collaboration]}, \emph{{The
  cosmic-ray electron spectrum measured with H.E.S.S.}},  Presentation at the
  35th International Cosmic Ray Conference, Busan, Korea, 2017,
  \href{https://indico.snu.ac.kr/indico/event/15/session/5/contribution/694}{https://indico.snu.ac.kr/indico/event/15/session/5/contribution/694}.

\bibitem{Genolini:2018jnu}
Y.~Genolini, \emph{{Theoretical interpretations of DAMPE first results: a
  critical review}},  Proceeding for the 53rd Rencontres de Moriond, 2018,
  \href{https://arxiv.org/abs/1806.06534}{{\ttfamily 1806.06534}}.

\bibitem{Galprop1}
A.~Strong, I.~Moskalenko, T.~Porter, G.~Johannesson, E.~Orlando, S.~Digel
  et~al., \emph{{GALPROP Version 54: Explanatory Supplement}}, 2011.

\bibitem{Strong:1998pw}
A.~W. Strong and I.~V. Moskalenko, \emph{{Propagation of cosmic-ray nucleons in
  the galaxy}}, \href{https://doi.org/10.1086/306470}{\emph{Astrophys. J.}
  {\bfseries 509} (1998) 212}
  [\href{https://arxiv.org/abs/astro-ph/9807150}{{\ttfamily
  astro-ph/9807150}}].

\bibitem{Evoli:2008dv}
C.~Evoli, D.~Gaggero, D.~Grasso and L.~Maccione, \emph{{Cosmic-Ray Nuclei,
  Antiprotons and Gamma-rays in the Galaxy: a New Diffusion Model}},
  \href{https://doi.org/10.1088/1475-7516/2008/10/018,
  10.1088/1475-7516/2016/04/E01}{\emph{JCAP} {\bfseries 0810} (2008) 018}
  [\href{https://arxiv.org/abs/0807.4730}{{\ttfamily 0807.4730}}].

\bibitem{Evoli:2016xgn}
C.~Evoli, D.~Gaggero, A.~Vittino, G.~Di~Bernardo, M.~Di~Mauro, A.~Ligorini
  et~al., \emph{{Cosmic-ray propagation with \texttt{DRAGON2}: I. numerical
  solver and astrophysical ingredients}},
  \href{https://doi.org/10.1088/1475-7516/2017/02/015}{\emph{JCAP} {\bfseries
  1702} (2017) 015} [\href{https://arxiv.org/abs/1607.07886}{{\ttfamily
  1607.07886}}].

\bibitem{Kissmann:2014sia}
R.~Kissmann, \emph{{PICARD: A novel code for the Galactic Cosmic Ray
  propagation problem}},
  \href{https://doi.org/10.1016/j.astropartphys.2014.02.002}{\emph{Astropart.
  Phys.} {\bfseries 55} (2014) 37}
  [\href{https://arxiv.org/abs/1401.4035}{{\ttfamily 1401.4035}}].

\bibitem{Maurin:2018rmm}
D.~Maurin, \emph{{USINE: semi-analytical models for Galactic cosmic-ray
  propagation}},  \href{https://arxiv.org/abs/1807.02968}{{\ttfamily
  1807.02968}}.

\bibitem{Eichler:1980hw}
D.~Eichler, \emph{{Basic inconsistencies in models of interstellar cosmic-ray
  acceleration}}, \href{https://doi.org/10.1086/157927}{\emph{Astrophys. J.}
  {\bfseries 237} (1980) 809}.

\bibitem{Cowsik:1980ApJ}
R.~Cowsik, \emph{{Comments on stochastic acceleration of cosmic rays}},
  {\emph{Astrophys. J.} {\bfseries 241} (1980) 1195}.

\bibitem{Fransson:1980ApJ}
C.~Fransson and R.~Epstein, \emph{{Acceleration and propagation of cosmic
  rays}}, {\emph{Astrophys. J.} {\bfseries 242} (1980) 411}.

\bibitem{1970ApJ...162L.181S}
C.~S. {Shen}, \emph{{Pulsars and Very High-Energy Cosmic-Ray Electrons}},
  \href{https://doi.org/10.1086/180650}{\emph{Astrophys.~J.~Lett.} {\bfseries
  162} (1970) L181}.

\bibitem{Atoian:1995ux}
A.~M. Atoian, F.~A. Aharonian and H.~J. Volk, \emph{{Electrons and positrons in
  the galactic cosmic rays}},
  \href{https://doi.org/10.1103/PhysRevD.52.3265}{\emph{Phys. Rev.} {\bfseries
  D52} (1995) 3265}.

\bibitem{Kobayashi:2003kp}
T.~Kobayashi, Y.~Komori, K.~Yoshida and J.~Nishimura, \emph{{The most likely
  sources of high energy cosmic-ray electrons in supernova remnants}},
  \href{https://doi.org/10.1086/380431}{\emph{Astrophys. J.} {\bfseries 601}
  (2004) 340} [\href{https://arxiv.org/abs/astro-ph/0308470}{{\ttfamily
  astro-ph/0308470}}].

\bibitem{DiMauro:2014iia}
M.~Di~Mauro, F.~Donato, N.~Fornengo, R.~Lineros and A.~Vittino,
  \emph{{Interpretation of AMS-02 electrons and positrons data}},
  \href{https://doi.org/10.1088/1475-7516/2014/04/006}{\emph{JCAP} {\bfseries
  1404} (2014) 006} [\href{https://arxiv.org/abs/1402.0321}{{\ttfamily
  1402.0321}}].

\bibitem{Green:2015isa}
D.~A. Green, \emph{{Constraints on the distribution of supernova remnants with
  Galactocentric radius}},
  \href{https://doi.org/10.1093/mnras/stv1885}{\emph{Mon. Not. Roy. Astron.
  Soc.} {\bfseries 454} (2015) 1517}
  [\href{https://arxiv.org/abs/1508.02931}{{\ttfamily 1508.02931}}].

\bibitem{Manchester:2004bp}
R.~N. Manchester, G.~B. Hobbs, A.~Teoh and M.~Hobbs, \emph{{The Australia
  Telescope National Facility pulsar catalogue}},
  \href{https://doi.org/10.1086/428488}{\emph{Astron. J.} {\bfseries 129}
  (2005) 1993} [\href{https://arxiv.org/abs/astro-ph/0412641}{{\ttfamily
  astro-ph/0412641}}].

\bibitem{Strong:2001qp}
A.~W. Strong and I.~V. Moskalenko, \emph{{A 3-D time dependent model for
  galactic cosmic rays and gamma-rays}},  in \emph{{27th International Cosmic
  Ray Conference (ICRC 2001) Hamburg, Germany, August 7-15, 2001}}, p.~1964,
  2001, \href{https://arxiv.org/abs/astro-ph/0106505}{{\ttfamily
  astro-ph/0106505}}.

\bibitem{Swordy:2003ds}
S.~Swordy, \emph{{Stochastic effects on the electron spectrum above TeV
  energies}},  in \emph{{Proceedings, 28th International Cosmic Ray Conference
  (ICRC 2003): Tsukuba, Japan, July 31-August 7, 2003}}, pp.~1989--1992, 2003,
  \href{http://www-rccn.icrr.u-tokyo.ac.jp/icrc2003/PROCEEDINGS/PDF/492.pdf}{http://www-rccn.icrr.u-tokyo.ac.jp/icrc2003/PROCEEDINGS/PDF/492.pdf}.

\bibitem{Lee:1979zz}
M.~A. {Lee}, \emph{{A statistical theory of cosmic ray propagation from
  discrete galactic sources}},
  \href{https://doi.org/10.1086/156970}{\emph{Astrophys. J.} {\bfseries 229}
  (1979) 424}.

\bibitem{2006AdSpR..37.1909P}
V.~S. {Ptuskin}, F.~C. {Jones}, E.~S. {Seo} and R.~{Sina}, \emph{{Effect of
  random nature of cosmic ray sources Supernova remnants on cosmic ray
  intensity fluctuations, anisotropy, and electron energy spectrum}},
  \href{https://doi.org/10.1016/j.asr.2005.08.036}{\emph{Advances in Space
  Research} {\bfseries 37} (2006) 1909}.

\bibitem{Blasi:2011fi}
P.~Blasi and E.~Amato, \emph{{Diffusive propagation of cosmic rays from
  supernova remnants in the Galaxy. I: spectrum and chemical composition}},
  \href{https://doi.org/10.1088/1475-7516/2012/01/010}{\emph{JCAP} {\bfseries
  1201} (2012) 010} [\href{https://arxiv.org/abs/1105.4521}{{\ttfamily
  1105.4521}}].

\bibitem{Mertsch:2010fn}
P.~Mertsch, \emph{{Cosmic ray electrons and positrons from discrete stochastic
  sources}}, \href{https://doi.org/10.1088/1475-7516/2011/02/031}{\emph{JCAP}
  {\bfseries 1102} (2011) 031}
  [\href{https://arxiv.org/abs/1012.0805}{{\ttfamily 1012.0805}}].

\bibitem{Bernard:2012wt}
G.~Bernard, T.~Delahaye, P.~Salati and R.~Taillet, \emph{{Variance of the
  Galactic nuclei cosmic ray flux}},
  \href{https://doi.org/10.1051/0004-6361/201219502}{\emph{Astron. Astrophys.}
  {\bfseries 544} (2012) A92}
  [\href{https://arxiv.org/abs/1204.6289}{{\ttfamily 1204.6289}}].

\bibitem{Gnedenko:1954}
B.~V. {Gendenko} and A.~N. Kolmogorov, \emph{{Limit Distributions for Sums of
  Independent Random Variables}}. Addison-Wesley, Cambridge, MA, USA, 1954.

\bibitem{Lagutin:1995rr}
A.~A. Lagutin and {\relax Yu}.~A. Nikulin, \emph{{Fluctuations and anisotropy
  of cosmic rays in the galaxy}}, {\emph{J. Exp. Theor. Phys.} {\bfseries 81}
  (1995) 825}.

\bibitem{Uchaikin:1999}
V.~V. {Uchaikin} and V.~M. Zolotarev, \emph{{Chance and Stability. Stable
  Distributions and Their Applications}}. VSP, Utrecht, 1999.

\bibitem{nolan:2010}
J.~P. Nolan, \emph{Stable Distributions - Models for Heavy Tailed Data}.
  Birkh\"auser, Boston, 2010.

\bibitem{Pohl:1998ug}
M.~Pohl and J.~A. Esposito, \emph{{Electron acceleration in SNR and diffuse
  gamma-rays above 1-GeV}},
  \href{https://doi.org/10.1086/306298}{\emph{Astrophys. J.} {\bfseries 507}
  (1998) 327} [\href{https://arxiv.org/abs/astro-ph/9806160}{{\ttfamily
  astro-ph/9806160}}].

\bibitem{Grasso:2009ma}
{\scshape Fermi-LAT} collaboration, D.~Grasso et~al., \emph{{On possible
  interpretations of the high energy electron-positron spectrum measured by the
  Fermi Large Area Telescope}},
  \href{https://doi.org/10.1016/j.astropartphys.2009.07.003}{\emph{Astropart.
  Phys.} {\bfseries 32} (2009) 140}
  [\href{https://arxiv.org/abs/0905.0636}{{\ttfamily 0905.0636}}].

\bibitem{Kawanaka:2009dk}
N.~Kawanaka, K.~Ioka and M.~M. Nojiri, \emph{{Cosmic-Ray Electron Excess from
  Pulsars is Spiky or Smooth?: Continuous and Multiple Electron/Positron
  injections}},
  \href{https://doi.org/10.1088/0004-637X/710/2/958}{\emph{Astrophys. J.}
  {\bfseries 710} (2010) 958}
  [\href{https://arxiv.org/abs/0903.3782}{{\ttfamily 0903.3782}}].

\bibitem{Blasi:2010de}
P.~Blasi and E.~Amato, \emph{{Positrons from pulsar winds}},  in
  \emph{{Proceedings, 1st Session of the Sant Cugat Forum on Astrophysics:
  High-Energy Emission from Pulsars and their Systems: Sant Cugat, Catalonia,
  Spain, April 12-16, 2010}}, pp.~623--641, 2011,
  \href{https://arxiv.org/abs/1007.4745}{{\ttfamily 1007.4745}},
  \href{https://doi.org/10.1007/978-3-642-17251-9_50}{DOI}.

\bibitem{Kashiyama:2010ui}
K.~Kashiyama, K.~Ioka and N.~Kawanaka, \emph{{White Dwarf Pulsars as Possible
  Cosmic Ray Electron-Positron Factories}},
  \href{https://doi.org/10.1103/PhysRevD.83.023002}{\emph{Phys. Rev.}
  {\bfseries D83} (2011) 023002}
  [\href{https://arxiv.org/abs/1009.1141}{{\ttfamily 1009.1141}}].

\bibitem{Ginzburg:1990sk}
V.~L. Ginzburg, (ed.~), V.~A. Dogiel, V.~S. Berezinsky, S.~V. Bulanov and V.~S.
  Ptuskin, \emph{{Astrophysics of cosmic rays}}. 1990.

\bibitem{1966ApJ...146..480J}
J.~R. {Jokipii}, \emph{{Cosmic-Ray Propagation. I. Charged Particles in a
  Random Magnetic Field}},
  \href{https://doi.org/10.1086/148912}{\emph{Astrophys.~J.} {\bfseries 146}
  (1966) 480}.

\bibitem{1966PhFl....9.2377K}
C.~F. {Kennel} and F.~{Engelmann}, \emph{{Velocity Space Diffusion from Weak
  Plasma Turbulence in a Magnetic Field}},
  \href{https://doi.org/10.1063/1.1761629}{\emph{Physics of Fluids} {\bfseries
  9} (1966) 2377}.

\bibitem{1967PhFl...10.2620H}
D.~E. {Hall} and P.~A. {Sturrock}, \emph{{Diffusion, Scattering, and
  Acceleration of Particles by Stochastic Electromagnetic Fields}},
  \href{https://doi.org/10.1063/1.1762084}{\emph{Physics of Fluids} {\bfseries
  10} (1967) 2620}.

\bibitem{1970ApJ...162.1049H}
K.~{Hasselmann} and G.~{Wibberenz}, \emph{{A Note on the Parallel Diffusion
  Coefficient}}, \href{https://doi.org/10.1086/150736}{\emph{Astrophys.~J.}
  {\bfseries 162} (1970) 1049}.

\bibitem{Blumenthal:1970gc}
G.~R. Blumenthal and R.~J. Gould, \emph{{Bremsstrahlung, synchrotron radiation,
  and compton scattering of high-energy electrons traversing dilute gases}},
  \href{https://doi.org/10.1103/RevModPhys.42.237}{\emph{Rev. Mod. Phys.}
  {\bfseries 42} (1970) 237}.

\bibitem{1959SvA.....3...22S}
S.~I. {Syrovatskii}, \emph{{The Distribution of Relativistic Electrons in the
  Galaxy and the Spectrum of Synchrotron Radio Emission.}}, {\emph{Soviet~Ast.}
  {\bfseries 3} (1959) 22}.

\bibitem{Ahlers:2009ae}
M.~Ahlers, P.~Mertsch and S.~Sarkar, \emph{{On cosmic ray acceleration in
  supernova remnants and the FERMI/PAMELA data}},
  \href{https://doi.org/10.1103/PhysRevD.80.123017}{\emph{Phys. Rev.}
  {\bfseries D80} (2009) 123017}
  [\href{https://arxiv.org/abs/0909.4060}{{\ttfamily 0909.4060}}].

\bibitem{Sklar:1959}
A.~Sklar, \emph{Fonctions de répartition à n dimensions et leurs marges},
  {\emph{Publ.\ Inst.\ Statist.\ Univ.\ Paris} (1959) 229}.

\bibitem{Aas:2009}
K.~Aas, C.~Czado, A.~Frigessi and H.~Bakken, \emph{Pair-copula constructions of
  multiple dependence},
  \href{https://doi.org/https://doi.org/10.1016/j.insmatheco.2007.02.001}{\emph{Insurance:
  Mathematics and Economics} {\bfseries 44} (2009) 182 }.

\bibitem{Joe:1996}
H.~Joe, \emph{Families of $m$-variate distributions with given margins and
  $m(m-1)/2$ bivariate dependence parameters}, vol.~Volume 28 of \emph{Lecture
  Notes--Monograph Series}, pp.~120--141.
\newblock Institute of Mathematical Statistics, 1996.
\newblock 10.1214/lnms/1215452614.

\bibitem{Bedford:2001}
T.~Bedford and R.~Cooke, \emph{Probability density decomposition for
  conditionally dependent random variables modeled by vines},
  \href{https://doi.org/10.1023/A:1016725902970}{\emph{Annals of Mathematics
  and Artificial Intelligence} {\bfseries 32} (2001) 245}.

\bibitem{Bedford:2002}
T.~Bedford and R.~M. Cooke, \emph{Vines--a new graphical model for dependent
  random variables}, \href{https://doi.org/10.1214/aos/1031689016}{\emph{Ann.
  Statist.} {\bfseries 30} (2002) 1031}.

\bibitem{Kurowicka:2005}
D.~Kurowicka and R.~M. Cooke, \emph{Sampling algorithms for generating joint
  uniform distributions using the vine-copula method},  3rd IASC World
  Conference on Computational Statistics \& Data Analysis, Limassol, Cyprus,
  2005.

\bibitem{Nagler:2016}
T.~Nagler and C.~Czado, \emph{{Evading the curse of dimensionality in
  nonparametric density estimation with simplified vine copulas}},
  \href{https://doi.org/https://doi.org/10.1016/j.jmva.2016.07.003}{\emph{Journal
  of Multivariate Analysis} {\bfseries 151} (2016) 69 }.

\bibitem{Vallee:2005}
J.~P. {Vall{\'e}e}, \emph{{The Spiral Arms and Interarm Separation of the Milky
  Way: An Updated Statistical Study}},
  \href{https://doi.org/10.1086/431744}{\emph{Astron. J.} {\bfseries 130}
  (2005) 569}.

\bibitem{Case:1998qg}
G.~L. Case and D.~Bhattacharya, \emph{{A new sigma-d relation and its
  application to the galactic supernova remnant distribution}},
  \href{https://doi.org/10.1086/306089}{\emph{Astrophys. J.} {\bfseries 504}
  (1998) 761} [\href{https://arxiv.org/abs/astro-ph/9807162}{{\ttfamily
  astro-ph/9807162}}].

\bibitem{1991ARA&A..29..363V}
S.~{van den Bergh} and G.~A. {Tammann}, \emph{{Galactic and extragalactic
  supernova rates}},
  \href{https://doi.org/10.1146/annurev.aa.29.090191.002051}{\emph{Ann. Rev.
  Astron. Astrophys.} {\bfseries 29} (1991) 363}.

\bibitem{Diehl:2006cf}
R.~Diehl et~al., \emph{{Radioactive Al-26 and massive stars in the galaxy}},
  \href{https://doi.org/10.1038/nature04364}{\emph{Nature} {\bfseries 439}
  (2006) 45} [\href{https://arxiv.org/abs/astro-ph/0601015}{{\ttfamily
  astro-ph/0601015}}].

\bibitem{Venter:2001}
G.~G. Venter, \emph{{Tails of Copulas}},  Presentation at ASTIN Colloquium,
  International Actuarial Association, Washington D.C., 2001.

\bibitem{Schirmacher:2008}
D.~Schirmacher and E.~Schirmacher, \emph{Multivariate dependence modeling using
  pair-copulas},  2008 ERM Symposium, Chicago, 2008.

\end{thebibliography}\endgroup

\end{document}